\documentclass[%
 reprint,
 amsmath,amssymb,
 aps,
]{revtex4-1}

\usepackage{graphicx}
\usepackage{dcolumn}
\usepackage{bm}

\begin{document}


\title{Boltzmann-type collision operators for Bogoliubov excitations of Bose-Einstein condensates: A unified framework}

\author{Minh-Binh Tran}%
 \email{minhbinht@mail.smu.edu}
\affiliation{Department of Mathematics, Southern Methodist University, Dallas, TX 75275, USA
}%
\author{Yves Pomeau}
 \email{yves.pomeau@gmail.com}
 \affiliation{LadHyX - Laboratoire d'hydrodynamique, Ladhyx, Ecole Polytechnique, Palaiseau, France.}


%

\date{\today}

\begin{abstract}
Starting from the  Bogoliubov diagonalization  for  the Hamiltonian of a weakly
interacting Bose gas under  the presence of  a Bose-Enstein Condensate (BEC),   we   derive  the kinetic equation for the  Bogoliubov excitations. Without dropping any of the commutators, we find three collisional processes.  One of  them describes the $1\leftrightarrow2$ interactions between the condensate and the excited atoms. The other two describe the $2\leftrightarrow2$ and $1\leftrightarrow3$ interactions between the excited atoms themselves.
\end{abstract}

\pacs{Valid PACS appear here}


\maketitle


\section{Introduction}
 Classical Boltzmann  kinetic theory is the way to connect the
macroscopic properties of gases of many particles to the fundamental
interaction by collisions between those particles. Shortly after the
establishment of quantum mechanics for bosons and fermions, Nordheim (cf.  \cite{Nordheim}) wrote the kinetic equations for dilute gases of quantum particles, that takes into account statistical effects linked to the possibility or not of overlaps of wave functions after a two body-interaction,  relying on the assumption that the strength of the interaction is small. The resulting Boltzmann-Nordheim kinetic theory is correct in principle for describing dilute quantum gases. However, it has to be changed for Bose gases at  low  temperatures to include a condensate. As being shown long ago by Bogoliubov \cite{bogoliubov1947theory}, the existence of this condensate changes the fundamental notion of what a particle is. Because of the interaction with the condensate, the notion of particles has to be modified by the one of quasi-particles, as guessed by Landau before. Even at equilibrium (see reference \cite{PomeauBinh}) a fully coherent theory
of  quasi particles is already a fairly non trivial issue. Of
course, it is even harder to derive a valid kinetic equation  for 
quasi particles. However, this kinetic theory is not too strongly
changed if the difference between particles and quasi particles is
restricted to a relatively small population of those particles, which
is the case if the kinetic energy per particle is still much larger
than its interaction energy with the condensate. If the
average kinetic energy per particle is of the order or less than interaction
with the condensate, one must take fully into account the Bogoliubov
renormalization of the particle energy, something that brings a lot of
terms in the kinetic equation. A major issue then is to derive what was
called by Reichl and collaborators \cite{ReichlGust:2012:CII} the $1\leftrightarrow3$ interactions  between the excited atoms, that
makes the main purpose of the present work. 

In the  work   \cite{KD1,KD2},    Kirkpatrick and Dorfman started to tackle the complex problem of writing the kinetic equation
for the gas of particles out of the condensate, coupled with those inside the
condensate, something that began with references \cite{akhiezer2013methods,PeletminskiiYatsenko:1968:FTB}. In this work, the authors derived a mean field kinetic equation  for a dilute condensed Bose gas that describes the relaxation in terms of ``collisions'' between  excitations.
The work of Kirkpatrick and Dorfman was then extended by Zaremba, Nikuni and Griffin \cite{ZarembaNikuniGriffin:1999:DOT}, where they introduced the full coupling system of a quantum Boltzmann equation for the density function of the normal fluid/thermal cloud and a Gross--Pitaevskii equation for the wavefunction of the BEC. Independently, the same model was also derived by Pomeau,  Brachet, M\'{e}tens and Rica in \cite{PomeauBrachetMetensRica}, using  the quantum BBGKY hierarchy argument. In a series of papers \cite{QK1,QK3,QK2},  Gardiner, Zoller and collaborators introduced a different model, which,  at the limits,  becomes the  model of Zaremba et al. and Pomeau et al.  We refer to \cite{GriffinNikuniZaremba:BCG:2009,PomeauBinh} for further discussions on the topic. In all of these kinetic equations,  there are two types of collisional processes:
 \begin{itemize}
 \item The $C_{12}$ collision operator describes the $1\leftrightarrow2$ interactions between the condensate and the excited atoms.
 \item The $C_{22}$ collision operator describes the $2\leftrightarrow2$ interactions between the excited atoms themselves.
 \end{itemize}
In \cite{ReichlGust:2012:CII},  Reichl and Gust  proposed the third, previously missing, collisional process, which takes into account $1{\leftrightarrow}3$ type collisions between the excitations, in addition to the $1{\leftrightarrow}2$ and $2{\leftrightarrow}2$ type collisions already known to occur. They called it  the collision operator ${C}_{31}$.

However, the derivation of the new collision operator ${C}_{31}$ is very complicated, since  the process  generates around $40000$ individual terms and one will need to do a combinatorics problem for all of them.  As a result, a    concise mathematical justification for the existence of the missing collision operator ${C}_{31}$ remains to be a  challenging open problem over the   years.

The aim of our work is to verify the validity of the collision operators $C_{12}, C_{22}, C_{31}$ by  a fairly simple framework. To this end, we focus only on the spatial homogeneous system. Our spatial homogeneous kinetic equation for the evolution of the density function $f(t,p)$ of the thermal cloud  takes the form

\begin{equation}
\label{KineticFinal}
\begin{aligned}
& \partial_t f(p) \ =  \ C_{12}[f](p) \  +  \  C_{22}[f](p)  \ + \ C_{31}[f](p),
\end{aligned}
\end{equation}
and the forms of $C_{12}$, $C_{22}$, $C_{31}$ are given explicitly below
\begin{equation}
\begin{aligned}\label{C12Discrete}
& C_{12}[f](t,p) \ =   \  4\pi {\frac{g^2n}{V}} \sum_{p_1,p_2,p_3\ne 0}(\delta(p-p_1)-\delta(p-p_2)\\
&-\delta(p-p_3)) \\
& \times \delta(\omega(p_1)-\omega(p_2)-\omega(p_3))(K_{1,2,3}^{1,2} )^2\delta(p_1-p_2-p_3)\\
& \times \Big[f(p_2)f(p_3)(f(p_1)+1)-f(p_1)(f(p_2)+1)(f(p_3)+1)\Big],
\end{aligned}
\end{equation}
\begin{equation}
\begin{aligned}\label{C22Discrete}
& C_{22}[f](t,p) \ =   \  \frac{g^2\pi}{V^2}  \sum_{p_1,p_2,p_3,p_4\ne 0}(\delta(p-p_1)+\delta(p-p_2)\\
& -\delta(p-p_3)-\delta(p-p_4))(K^{2,2}_{1,2,3,4})^2\\
&\times \delta(p_1+p_2-p_3-p_4)\delta(\omega(p_1)+\omega(p_2)-\omega(p_3)-\omega(p_4))\\
&\times \Big[f(p_3)f(p_4)(f(p_2)+1)(f(p_1)+1)\\
& -f(p_1)f(p_2)(f(p_3)+1)(f(p_4)+1)\Big],
\end{aligned}
\end{equation}
and
\begin{equation}
\begin{aligned}\label{C31Discrete}
& C_{31}[f](t,p)  = \ {\frac{3g^2\pi}{V}} \sum_{p_1,p_2,p_3,p_4\ne 0}(\delta(p-p_1)-\delta(p-p_2)\\
&-\delta(p-p_3)-\delta(p-p_4))\\
&\times (K^{3,1}_{1,2,3,4})^2\delta(p_1-p_2-p_3-p_4)\\
&\times\delta(\omega(p_1)-\omega(p_2)-\omega(p_3)-\omega(p_4))\\
&\times \Big[f(p_3)f(p_4)f(p_2)(f(p_1)+1)\\
&-f(p_1)(f(p_2)+1)(f(p_3)+1)(f(p_4)+1)\Big],
\end{aligned}
\end{equation}
in which $n$ is the density of the condensate, $t\in\mathbb{R}_+$ is the time variable, $p\in\mathbb{R}^d\backslash\{O\}$ is the $d$-dimensional non-zero momentum variable, $V$ is proportional to the volume of the periodic box  $\left[-\frac{L}{2},\frac{L}{2}\right]^d$, $\omega$ is the Bogoliubov dispersion relation defined in \eqref{BogDispersion}, $g$ is the interacting constant. We have normalized the Plank constant to be $1$.

In the above collision operators, the kernels are defined as follows
\begin{equation}
\label{KernelC12}\begin{aligned}
K_{1,2,3}^{1,2}  = & \ u_{p_1}u_{p_2}u_{p_3}-v_{p_1}v_{p_2}v_{p_3}-u_{p_1}u_{p_2}v_{p_3}\\
&+v_{p_1}v_{p_2}u_{p_3}-u_{p_1}v_{p_2}u_{p_3}+v_{p_1}u_{p_2}v_{p_3},\end{aligned}
\end{equation}
\begin{equation}
\label{KernelC22}\begin{aligned}
K^{2,2}_{1,2,3,4} \ = & \ u_{p_1}u_{p_2}u_{p_3}u_{p_4}+u_{p_1}v_{p_2}u_{p_3}v_{p_4}+u_{p_1}v_{p_2}v_{p_3}u_{p_4}\\
& +v_{p_1}u_{p_2}v_{p_3}u_{p_4}+v_{p_1}u_{p_2}u_{p_3}v_{p_4}+v_{p_1}v_{p_2}v_{p_3}v_{p_4},\end{aligned}
\end{equation}
and
\begin{equation}
\label{KernelC31}\begin{aligned}
K^{3,1}_{1,2,3,4} \ = & \ 2\Big[u_{p_1}u_{p_2}v_{p_3}u_{p_4}+v_{p_1}v_{p_2}u_{p_3}v_{p_4}\Big],\end{aligned}
\end{equation}
with $u_p$ and $v_p$ being defined later in \eqref{BogConstant}.

To derive \eqref{KineticFinal}, we start with the  Bogoliubov diagonalization process  for  the Hamiltonian of a weakly
interacting Bose gas under  the presence of  a BEC, then focus on the derivation of the kinetic equation for  the Bogoliubov excitations. In this process, we compute all of the commutators of the Bogoliubov excitations  and do not drop any of them. We discover special mathematical structures of the commutators that allow us to reduce significantly the number of terms and the amount of computations.  Especially, the computations of $C_{31}$ reduce from $40000$ to only around $30$ terms. Therefore, the combinatorics problem can simply  be done and checked  by hand.

 Moreover, our framework provides a unified  point of view  for the different models, as it gives a simple explanation for the origins of the different collision operators based on the Bogoliubov diagonalisation. To see this, we note that after the Bogoliubov  transformation, the nonlinearity $\hat{a}^\dagger \hat{a}^\dagger \hat{a}\hat{a}$ of the Hamiltonian of the quantum system contains several types of nonlinearities including the following 3 special ones: (i) $\hat{b}^\dagger \hat{b}^\dagger \hat{b}$ and $\hat{b}^\dagger\hat{b} \hat{b};$  (ii) $\hat{b}^\dagger \hat{b}^\dagger \hat{b}\hat{b};$ (iii) $\hat{b}^\dagger \hat{b}^\dagger \hat{b}^\dagger \hat{b}$ and $\hat{b}^\dagger\hat{b} \hat{b}\hat{b}$; where  $ \hat{a}^\dagger, \hat{a}$   are  bosonic creation and annihilation operators and $ \hat{b}^\dagger, \hat{b}$ are their Bogoliubov transformations.  The 3 types of collision operators then appear naturally as combinations of commutators of each type as follows.
\begin{itemize}
\item The $C_{12}$ collision operator arises from commutators of the type $[\hat{b}^\dagger \hat{b}^\dagger \hat{b},[\hat{b}^\dagger \hat{b},\hat{b}^\dagger\hat{b} \hat{b}]]$ and $[\hat{b}^\dagger\hat{b} \hat{b} ,[\hat{b}^\dagger \hat{b},\hat{b}^\dagger \hat{b}^\dagger \hat{b}]]$.
\item The $C_{22}$ (Boltzmann-Nordheim/Uehling-Ulenbeck) collision operator arises from  commutators of the type $[\hat{b}^\dagger \hat{b}^\dagger \hat{b}\hat{b},[\hat{b}^\dagger \hat{b},\hat{b}^\dagger \hat{b}^\dagger \hat{b}\hat{b}]]$.
 \item The $C_{31}$ collision operator arises from  commutators of the types $[\hat{b}^\dagger \hat{b}^\dagger \hat{b}^\dagger \hat{b},[\hat{b}^\dagger \hat{b},\hat{b}^\dagger\hat{b} \hat{b}\hat{b}]]$ and $[\hat{b}^\dagger \hat{b}\hat{b}\hat{b},[\hat{b}^\dagger \hat{b},\hat{b}^\dagger \hat{b}^\dagger \hat{b}^\dagger \hat{b}]]$.
\end{itemize}
The above argument provides a concise mathematical   confirmation of  the existence of $C_{31}$. For the experimental confirmations  of $C_{31}$, we refer the readers to \cite{PomeauBinh,reichl2013transport}.

To conclude the introductory section, we remark that, when the temperature of the system is lower but closed to the Bose-Einstein condensation transition temperature, the Bogoliubov dispersion relation can be approximated by the Hatree-Fock energy. In this case, $u_p\backsim 1$ and $v_p\backsim 0$. As a result,   the kernel $ K_{1,2,3}^{1,2}\backsim 1$ and  the kernel $K^{2,2}_{1,2,3,4}\backsim 1$. On the other hand, the kernel  $K^{3,1}_{1,2,3,4}\backsim 0$. As a result, in this temperature regime, the two collision operators $C_{12}$ and $C_{22}$ dominate the collisional processes. The contribution of third collision operator $C_{31}$ becomes significant when both $u_p$ and $v_p$ are large, corresponding to lower temperature regimes. 

\section{The quantum system and the Bogoliubov transformation}
To begin our quantum description, since we are studying an interacting many body quantum system, in which, dealing with the wavefunction for each individual particle becomes  cumbersome, we introduce  the boson field operator $\hat\Psi(x)$, and its conjugate $\hat\Psi^\dagger(x)$. These operators satisfy the the commutation
relation
\begin{equation}
\begin{aligned}
& [\hat\Psi(x),\hat\Psi(x')] \  =\ [\hat\Psi^\dagger(x),\hat\Psi^\dagger(x')]\ =\ 0, \\
&  \mbox{ and  } [\hat\Psi(x),\hat\Psi^\dagger(x')] \ = \ \delta(x-x').
\end{aligned}
\end{equation}
The Hamiltonian of the system is now written
\begin{equation}
\label{NLS}\begin{aligned}
\hat{H} \ = & \ \int_{\mathbb{T}_L^d}\mathrm{d}x\hat\Psi^\dagger(x)\Big[-\frac{\hbar^2}{2m}\nabla^2\\
& \ +  \ U(x) \ + \ \frac{1}{2}\int_{\mathbb{T}_L^d}\mathrm{d}x'\hat\Psi^\dagger (x) \mathcal{V}(x,x')\hat\Psi(x') \Big]\hat\Psi(x),\end{aligned}
\end{equation}
where $\mathbb{T}_L^d$ is the $d$-dimensional periodic torus $\left[-\frac{L}{2},\frac{L}{2}\right]^d$; $\hbar$ is the Planck constant; $m$ is the mass of the particle; $U$ is a externally applied potential; $ \mathcal{V}(x,x')$ is the interaction potential between two
particles at locations $x$, $x'$. To simplify our settings, we will not discuss particles in an external trapping potential, and set $U=0$. We also take $\mathcal{V}(x,x') =g\delta(x-x')$, where $g$ is the interacting constant. Inserting these two forms for the external and interaction potentials into \eqref{NLS}, we find
\begin{equation}
\label{NLS1}
\hat{H}  =  \int_{\mathbb{T}_L^d}\mathrm{d}x\left[-\frac{\hbar^2}{2m}\hat\Psi^\dagger(x)\nabla^2\hat\Psi(x) +  \frac{g}{2}\hat\Psi^\dagger (x)\hat\Psi^\dagger (x) \hat\Psi(x) \hat\Psi(x) \right].
\end{equation}
Writing the wave function $\Psi$ in terms of annihilation and creation operators, we obtain
\begin{equation}
\label{Fourier1}
\hat{a}_p(t) \  = \ \frac{1}{(2\pi L)^\frac{d}{2}}\int_{\mathbb{T}_L^d}\mathrm{d}x e^{-ipx}\hat\Psi(t,x),
\end{equation} 
and
\begin{equation}
\label{Fourier2}
\hat\Psi(t,x) \ = \ \frac{1}{(2\pi L )^\frac{d}{2}}\sum_{p\in\mathbb{Z}_L^d}e^{ipx}\hat{a}_p(t),
\end{equation}
where $\mathbb{Z}_L^d = (\mathbb{Z}/L)^d $.  For the sake of simplicity, we employ the shorthand notations
\begin{equation}
\label{Shorthand}
\int_{\mathbb{T}_L^d} \ = \ \int ,\ \ \  \ \ \ \sum_{p\in\mathbb{Z}_L^d} \  = \ \sum_{p} \ \ \, \mbox{ and } V=(2\pi L)^d.
\end{equation}
The annihilation and creation operators $\hat{a}_p$ and $\hat{a}_p^\dagger$ then satisfy the commutation relations
\begin{equation}
\label{BosonCommutation}
[\hat{a}_p,\hat{a}_{p'}]=[\hat{a}^\dagger_p,\hat{a}^\dagger_{p'}]=0, \mbox{ and } [\hat{a}_p,\hat{a}_{p'}^\dagger] = \delta(p-p').
\end{equation}
The Hamiltonian of the above system  is then
\begin{equation}
\label{Hamiltonian}
{H} =  \sum_{p}\epsilon_p \hat{a}_p^\dagger \hat{a}_p +  \frac{g}{2V} \sum_{p,p_1,p_2,p_3}\delta(p+p_1-p_2-p_3)\hat{a}_p^\dagger  \hat{a}_{p_1}^\dagger \hat{a}_{p_2} \hat{a}_{p_3},
\end{equation}
in which $ \epsilon_p=\frac{p^2}{2m}$ and the function $\delta(p+p_1-p_2-p_3)$ means that we sum over $p, p_1,p_2,p_3\in \mathbb{Z}_L^d$ such that $p+p_1=p_2+p_3$. We set $\hbar=1$, for the sake of simplicity. 

The Bose-Einstein condensation occurs when a large number of cold bosons enter the same quantum state having zero momentum. According to the Bogoliubov theory \cite{bogoliubov1947theory}, since the lowest energy state is occupied by macroscopic number of particles in the condensate, one can neglect the quantum fluctuation of this state and replace its annihilation operator with a $c$-number $\sqrt{N}$, with $N$ being the number of condensate atoms
\begin{equation}
\label{hata0}\hat{a}_0=\sqrt{N}. 
\end{equation}

We now split $\hat{a}_0$ and $\hat{a}_p$ ($p\ne 0$) and decompose the Hamiltonian $\hat{H}$ as 
\begin{equation}\label{HamiltonianExcitations}
\begin{aligned}
\hat{H} \ = \ \hat{H}_1 \ + \ \hat{H}_2 \ + \ \hat{H}_3, 
\end{aligned}
\end{equation}
with
\begin{equation}\label{HamiltonianExcitations1}
\begin{aligned}
\hat{H}_1 \ & = \  \sum_{p\ne 0} \epsilon_p \hat{a}_p^\dagger \hat{a}_p \ + \ \frac{g}{2V}\hat{a}_0^\dagger \hat{a}_0^\dagger \hat{a}_0 \hat{a}_0 \\
&  + \ \frac{g}{2V}\sum_{p\ne 0}\Big[4\hat{a}_0^\dagger \hat{a}_p^\dagger \hat{a}_0 \hat{a}_p \ + \ \hat{a}^\dagger_p \hat{a}^{\dagger}_{-p}\hat{a}_0\hat{a}_0 \ + \ \hat{a}_0^\dagger \hat{a}_0^\dagger \hat{a}_p \hat{a}_{-p}\Big],
\end{aligned}
\end{equation}
\begin{equation}\label{HamiltonianExcitations2}
\begin{aligned}
\hat{H}_2  \ = \  & \frac{g\sqrt{N}}{V}\sum_{p_1,p_2,p_3\ne 0}\Big[\delta(p_1-p_2-p_3)\hat{a}_{p_1}^\dagger \hat{a}_{p_2} \hat{a}_{p_3}\\
& +\delta(p_1+p_2-p_3)\hat{a}_{p_1}^\dagger \hat{a}_{p_2}^\dagger \hat{a}_{p_3}\Big],
\end{aligned}
\end{equation}
\begin{equation}\label{HamiltonianExcitations3}
\begin{aligned}
\hat{H}_3 \ = \  \frac{g}{2V}\sum_{p_1,p_2,p_3,p_4\ne 0}\delta(p_1+p_2-p_3-p_4)\hat{a}_{p_1}^\dagger \hat{a}_{p_2}^\dagger \hat{a}_{p_3}\hat{a}_{p_4}.
\end{aligned}
\end{equation}
Defining the  density $n=\frac{N}{V}$, we then find
\begin{equation}\label{HamiltonianExcitations1b}
\begin{aligned}
\hat{H}_1 \ = & \ \sum_{p\ne 0} \epsilon_p \hat{a}_p^\dagger \hat{a}_p \ + \ \frac{gnN}{2} \\
&  + \ \frac{gn}{2}\sum_{p\ne 0}\Big[2 \hat{a}_p^\dagger  \hat{a}_p \ + \ \hat{a}^\dagger_p \hat{a}^{\dagger}_{-p} \ + \  \hat{a}_p \hat{a}_{-p}\Big],
\end{aligned}
\end{equation}
which can be diagonalized using the Bogoliubov transformation
\begin{equation}
\label{BogoliubovDiagonalization}
\hat{a}_p \ = \ u_p \hat{b}_p \ - \ v_p \hat{b}_{-p}^\dagger, \ \ \ \hat{a}^\dagger_p \ = \ u_p \hat{b}^\dagger_p \ - \ v_p\hat{b}_{-p},
\end{equation}
with
\begin{equation}
\label{BogConstant}
u_p, v_p \ = \ \left(\frac{\epsilon_p +gn}{2\omega_p}\pm\frac12\right)^\frac12,
\end{equation}
where $\omega_p$ is the Bogoliubov dispersion relation 
\begin{equation}
\label{BogDispersion}
\omega_p \ = \ \left[\frac{gn}{m}p^2 \ + \ \left(\frac{p^2}{2m}\right)^2\right]^\frac12.
\end{equation}
After being diagonalized, $H_1$ takes the form
\begin{equation}
\label{H1Diagonalized}
\hat{H}_1 \ = \ \sum_{p\ne 0}\omega_p \hat{b}^\dagger \hat{b}_p \ + \ E_0,
\end{equation}
with
\begin{equation}
\label{E0}
E_0 \ = \ \frac{gnN}{2} \ + \frac12\sum_{p\ne 0}\left[\omega_p \ - \ gn \ - \ \frac{p^2}{2m} \ + \ \frac{m(gn)^2}{p^2}\right].
\end{equation}
\section{New forms of $\hat{H}_2$ and $\hat{H}_3$}
\subsection{New form of $\hat{H}_2$}
By the calculations to be detailed in Appendix A, we arrive at the following form of $\hat{H}_2$ in terms of the new operators $\hat{b}$ and $\hat{b}^\dagger$
\begin{equation}
\label{H2Eq5}
\hat{H}_2 \ = \ \hat{H}_{1,2} \ + \ \hat{H}_{3,0},
\end{equation}
where
\begin{equation}
\label{H2eqMain}\begin{aligned}
\hat{H}_{1,2} \ = & \ g\sqrt{\frac{n}{V}}\sum_{p_1,p_2,p_3\ne 0}\delta(p_1-p_2-p_3)K_{1,2,3}^{1,2}\\
&\times (\hat{b}_{p_1}^\dagger \hat{b}_{p_2}\hat{b}_{p_3} + \hat{b}^\dagger_{p_3} \hat{b}^\dagger_{p_2}\hat{b}_{p_1}),\\
K_{1,2,3}^{1,2}  = & u_{p_1}u_{p_2}u_{p_3}-v_{p_1}v_{p_2}v_{p_3}-u_{p_1}u_{p_2}v_{p_3}\\
&+v_{p_1}v_{p_2}u_{p_3}-u_{p_1}v_{p_2}u_{p_3}+v_{p_1}u_{p_2}v_{p_3},\end{aligned}
\end{equation}
and
\begin{equation}
\label{H2eqMain}\begin{aligned}
\hat{H}_{3,0} \ = & \ g\sqrt{\frac{n}{V}}\sum_{p_1,p_2,p_3\ne 0}\delta(p_1+p_2+p_3)\\
&\times\Big[K^{3,0}_{1,2,3}(\hat{b}_{p_3}^\dagger \hat{b}_{p_2}^\dagger \hat{b}_{p_1}^\dagger + \hat{b}_{p_1} \hat{b}_{p_2}\hat{b}_{p_3})\Big],\\
K^{3,0}_{1,2,3} \ = & \ u_{p_1}v_{p_2}v_{p_3}-v_{p_1}u_{p_2}u_{p_3}.\end{aligned}
\end{equation}

The Hamiltonian $\hat{H}_{1,2}$ contains strings of annihilation and creation operators of the types  $\hat{b}^\dagger \hat{b}\hat{b}$ and $\hat{b}^\dagger \hat{b}^\dagger \hat{b}$, that indicate the processes of one/two Bogoliubov excitations being created while two/one Bogoliubov excitations being annihilated.  On the other hand, the Hamiltonian $\hat{H}_{3,0}$ contains strings of annihilation and creation operators of the types  $\hat{b}\hat{b}\hat{b}$ and $\hat{b}^\dagger \hat{b}^\dagger \hat{b}^\dagger$, representing the process of the creation or annihilation of three excitations simultaneously. From a physical point of view, we can see that $\hat{H}_{3,0}$ does not contribute to the collision integrals, while the main contribution comes from  $\hat{H}_{1,2}$. We will show later in Section \ref{Sec:QuantumLiouville}, by explicit computations, that this is indeed the case. 
\subsection{New form of $\hat{H}_3$}
Similarly, we also find a new form for $\hat{H}_3$. The details of this computation will be given in Appendix B
\begin{equation}
\label{H3NewForm}
\hat{H}_3 \ = \ \hat{H}_{2,2} \ + \ \hat{H}_{1,1}\ + \ \hat{H}_{2,2}' \ + \  \ \hat{H}_{3,1} \ + \ \hat{H}_{3,1}' \ + \ \hat{H}_{4,0},
\end{equation}
where
\begin{equation}
\label{H22}\begin{aligned}
\hat{H}_{2,2} \ = & \ \frac{g}{2V}\sum_{p_1,p_2,p_3,p_4\ne 0}\delta(p_1+p_2-p_3-p_4)K^{2,2}_{1,2,3,4}\\
&\times \hat{b}_{p_1}^\dagger \hat{b}_{p_2}^\dagger \hat{b}_{p_3}\hat{b}_{p_4},\\
K^{2,2}_{1,2,3,4} \ = & \ u_{p_1}u_{p_2}u_{p_3}u_{p_4}+u_{p_1}v_{p_2}u_{p_3}v_{p_4}+u_{p_1}v_{p_2}v_{p_3}u_{p_4}\\
& +v_{p_1}u_{p_2}v_{p_3}u_{p_4}+v_{p_1}u_{p_2}u_{p_3}v_{p_4}+v_{p_1}v_{p_2}v_{p_3}v_{p_4},\end{aligned}
\end{equation}
\begin{equation}
\label{HL}
\hat{H}_{1,1} =\frac{g}{2V} \sum_{p_1,p_2\ne 0}K^{1,1}_{1,2}\hat{b}^\dagger_{p_1} \hat{b}_{p_1},
\end{equation}
\begin{equation}
\label{HL}
K^{1,1}_{1,2} =4v_{p_1}^2v_{p_2}^2+4 u_{p_1}^2v_{p_2}^2+4u_{p_1}v_{p_1}u_{p_2}v_{p_2},
\end{equation}
\begin{equation}
\label{H22primeprime}
\hat{H}_{2,2}' =\frac{g}{2V} \sum_{p_1,p_2\ne 0}\Big[u_{p_1}v_{p_1}u_{p_2}v_{p_2} \ + \ 2 v_{p_1}^2v_{p_2}^2\Big],
\end{equation}
\begin{equation}
\label{H31}\begin{aligned}
\hat{H}_{3,1} \ = & \ \frac{g}{2V}\sum_{p_1,p_2,p_3,p_4\ne 0}\delta(p_1-p_2-p_3-p_4)K^{3,1}_{1,2,3,4}\\
&\times \Big[\hat{b}_{p_1}^\dagger \hat{b}_{p_2} \hat{b}_{p_3}\hat{b}_{p_4}+\hat{b}_{p_4}^\dagger \hat{b}_{p_3}^\dagger \hat{b}_{p_2}^\dagger \hat{b}_{p_1}\Big],\\
K^{3,1}_{1,2,3,4} \ = & \ 2\Big[u_{p_1}u_{p_2}v_{p_3}u_{p_4}+v_{p_1}v_{p_2}u_{p_3}v_{p_4}\Big],\end{aligned}
\end{equation}
\begin{equation}
\label{H31prime}\begin{aligned}
\hat{H}'_{3,1}   = \ &  \frac{g}{2V}\sum_{p_1,p_2\ne 0}\Big[\hat{b}_{p_1} \hat{b}_{-p_1}K^{2,0}_{1,2} \ 
 +  \ \hat{b}_{p_1}^\dagger \hat{b}_{-p_1}^\dagger K^{2,0}_{1,2}\Big],\end{aligned}
\end{equation}
\begin{equation}
\label{KL2}
K^{2,0}_{1,2} \ = \ u_{p_1}^2 u_{p_2}v_{p_2}+v_{p_1}^2 u_{p_2}v_{p_2}+ 4u_{p_1}v_{p_1}v_{p_2}^2
\end{equation}
and
\begin{equation}
\label{H1prime}\begin{aligned}
\hat{H}_{4,0}\ =& \ \frac{g}{2V}\sum_{p_1,p_2,p_3,p_4\ne 0}\delta(p_1+p_2+p_3+p_4)\\
&\times K^{4,0}_{1,2,3,4}  \Big[\hat{b}_{p_1}^\dagger \hat{b}_{p_2}^\dagger \hat{b}_{p_3}^\dagger \hat{b}_{p_4}^\dagger + \hat{b}_{p_1} \hat{b}_{p_2}\hat{b}_{p_3}\hat{b}_{p_4}\Big],\end{aligned}
\end{equation}
with\begin{equation}
\label{K40}
K^{4,0}_{1,2,3,4}   \  =  \  u_{p_1}u_{p_2}v_{p_3}v_{p_4}.
\end{equation}
We remark that the Hamiltonian $\hat{H}_{2,2}$ contains strings of annihilation and creation operators of the types  $\hat{b}^\dagger \hat{b}^\dagger \hat{b} \hat{b}$,  indicating the processes of two Bogoliubov excitations being created while two Bogoliubov excitations being annihilated. Similarly, the Hamiltonian $\hat{H}_{3,1}$ represents the processes of three/one Bogoliubov excitations being created while one/tree Bogoliubov excitations being annihilated.  From a physical point of view, the main contribution to the collision integrals comes from  $\hat{H}_{2,2}$, $\hat{H}_{3,1}$ since  the effects of $\hat{H}_{4,0}, \hat{H}_{2,2}', \hat{H}_{3,1}'$, $\hat{H}_{1,1} $ are similar with that  of the Hamiltonian $\hat{H}_{3,0}$ discussed above and can be ignored. In Section \ref{Sec:QuantumLiouville}, this prediction will be shown by a more precise mathematical argument. 

\section{The quantum Liouville equation and assumptions}\label{Sec:QuantumLiouville}
The full state of the system is described by the full density matrix $\hat{\rho}$(t) which obeys the quantum Liouville
equation
\begin{equation}
\label{Liouville}
\partial_t \hat\rho \ = \ -{i}[\hat{H},\hat\rho].
\end{equation}

In order to derive  the quantum kinetic equation, there are two key points:

\begin{itemize}
\item First, due to the uncertainty principle, we cannot specify exactly the number of particles at positions and momenta. We can only describe the  number distribution of particles  in a quantum state. 
 As a consequence, the average number of quantum particles in quantum states with wave vectors  can be considered to be analogous of the average number of classical particle with momenta.
\item Second, in order to derive the quantum Boltzmann equation, we impose the Bogoliubov assumption that for a system that is out of equilibrium, the relaxation to equilibrium can occur in many different stages in which the stages' timescales are totally different from one stage to another.  During the relaxation process, in each successive stage, the set of relevant parameters (expectation values and mean fields) used to describe the evolution  is reduced. The Bogoliubov assumption is very similar to the {\it molecular chaos assumption} which implies  that the system can be described by a reduced number of parameters, for example, the single particle phase space distribution function.   
\end{itemize}
Employing the standard elimination process (cf. \cite{akhiezer2013methods,PeletminskiiYatsenko:1968:FTB}), we get the following spatial homogeneous equation for the single particle phase space distribution function $f(p)=\langle \hat{b}_p^\dagger \hat{b}_p \rangle=\mathrm{Tr}\Big(\hat{\rho}_0\hat{b}_p^\dagger  \hat{b}_p\Big)$
\begin{equation}
\label{Liouville1}
\partial_t f \ = \ \int_{-\infty}^0\mathrm{ds}\mathrm{Tr}\Big(\hat\rho_0[\hat{H},[f,\tilde{\hat{H}}(s)]]\Big),
\end{equation}
where $\tilde{\hat{H}}$ has exactly the same form with $H(s)$, except that all the operator $\hat{b}$, $\hat{b}^\dagger$ are replaced by $e^{is\omega} \hat{b}$ and $e^{-is\omega}\hat{b}^\dagger$. And following  \cite{akhiezer2013methods,PeletminskiiYatsenko:1968:FTB}
\begin{equation}
\label{rho0}\hat\rho_0 \ = \ \exp\left(-\sum_p \xi_p \hat{b}_p^\dagger \hat{b}_p- \Omega\right),
\end{equation}
with $\Omega=\log\mathrm{Tr}\left(\exp\left(-\sum_p \xi_p \hat{b}_p^\dagger b_p\right)\right)$.

We set $E_0=0$ and $H_{2,2}'=0$, since they are constants, and   approximate the right hand side of \eqref{Liouville1} as 
\begin{equation}\label{Liouville2}\begin{aligned}
& \int_{-\infty}^0\mathrm{ds}\mathrm{Tr}\Big(\hat\rho_0[\hat{H}(t),[f(s),\tilde{\hat{H}}(s)]] \Big)\\
\ \approx \ & \mathfrak{L}_{1,2} \ + \ \mathfrak{L}_{3,0} \ + \ \mathfrak{L}_{2,2} \ + \ \mathfrak{L}_{3,1} \ + \ \mathfrak{L}_{4,0} \ + \ \mathfrak{L}_{1,1} \ + \ \mathfrak{L}_{2,0}.\end{aligned}
\end{equation}
Notice  that in \cite{KD2},  the authors used  an  equivalent process, but only  kept $\mathfrak{L}_{1,2}$ to get the  $1\leftrightarrow2$  collision operator for the low temperature  regime, while  we keep all of the $7$  terms.

The forms of $\mathfrak{L}_{1,2}$, $\mathfrak{L}_{3,0}$, $\mathfrak{L}_{2,2} $, $\mathfrak{L}_{3,1}$, $\mathfrak{L}_{4,0}$, $\mathfrak{L}_{1,1}$, $\mathfrak{L}_{2,0}$ are computed as follows.

{\it The form of $\mathfrak{L}_{1,2}$.} This quantity comes from $\hat{H}_{1,2}$
\begin{equation}
\label{L12}
\begin{aligned}
\mathfrak{L}_{1,2} \ = & \ \int_{-\infty}^0\mathrm{ds}\mathrm{Tr}\Big(\hat\rho_0[\hat{H}_{1,2},[f,\tilde{\hat{H}}_{1,2}(s)]]\Big),
\end{aligned}
\end{equation}
where $\tilde{\hat{H}}_{1,2}$ has exactly the same form with ${\hat{H}}_{1,2}(s)$, except that all the operator $\hat{b}$, $\hat{b}^\dagger$ are replaced by $e^{is\omega} \hat{b}$ and $e^{-is\omega}\hat{b}^\dagger$.
Adapting the  procedure in \cite{akhiezer2013methods,KD2,PeletminskiiYatsenko:1968:FTB}, we write 
\begin{equation}
\label{L12a}
\begin{aligned}
\mathfrak{L}_{1,2} 
\ \approx & \  {\frac{g^2n}{V}}\sum_{p_1,p_2,p_3,p_1',p_2',p_3'\ne 0}\int_{-\infty}^0\mathrm{d}se^{is(\omega(p_1)-\omega(p_2)-\omega(p_3))}\\
&\times K^{1,2}_{1,2,3}K^{1,2}_{1',2',3'}\delta(p_1-p_2-p_3)\delta(p_1'-p_2'-p_3')\\
& \times \mathrm{Tr}\Big(\hat\rho_0\Big[[\hat{b}^\dagger_{p_3'}\hat{b}^\dagger_{p_2'}\hat{b}_{p_1'},[\hat{b}^\dagger_p\hat{b}_p,\hat{b}^\dagger_{p_1}\hat{b}_{p_2}\hat{b}_{p_3}]]\\
&+ [\hat{b}_{p_1'}^\dagger \hat{b}_{p_2'}\hat{b}_{p_3'},[\hat{b}_p^\dagger \hat{b}_p,\hat{b}_{p_3}^\dagger \hat{b}_{p_2}^\dagger \hat{b}_{p_1}]]\Big]\Big).
\end{aligned}
\end{equation}
 $K^{1,2}_{1',2',3'}$ has the same formulation with $K^{1,2}_{1,2,3}$, in which $p_1,p_2,p_3$ are replaced by $p_1',p_2',p_3'$.
We approximate (cf. \cite{akhiezer2013methods,PeletminskiiYatsenko:1968:FTB})
\begin{equation}
\label{Delta}
\int_{-\infty}^0\mathrm{d}se^{is(\omega(p_1)-\omega(p_2)-\omega(p_3))} \ \approx \ \pi\delta(\omega(p_1)-\omega(p_2)-\omega(p_3)),
\end{equation}
and write
\begin{equation}
\label{L30}
\begin{aligned}
&\mathfrak{L}_{1,2}
 \approx  \frac{g^2n\pi}{V} \sum_{p_1,p_2,p_3,p_1',p_2',p_3'\ne 0}K^{1,2}_{1,2,3}K^{1,2}_{1',2',3'}\\
& \times\delta(\omega(p_1)-\omega(p_2)-\omega(p_3)) \delta(p_1-p_2-p_3)\delta(p_1'-p_2'-p_3')\\
&\times \mathrm{Tr}\Big(\hat\rho_0\Big[[\hat{b}^\dagger_{p_3'}\hat{b}^\dagger_{p_2'}\hat{b}_{p_1'},[\hat{b}^\dagger_p\hat{b}_p,\hat{b}^\dagger_{p_1}\hat{b}_{p_2}\hat{b}_{p_3}]]\\
&+ [\hat{b}_{p_1'}^\dagger \hat{b}_{p_2'}\hat{b}_{p_3'},[\hat{b}_p^\dagger \hat{b}_p,\hat{b}_{p_3}^\dagger \hat{b}_{p_2}^\dagger \hat{b}_{p_1}]]\Big]\Big).
\end{aligned}
\end{equation}

{\it The form of $\mathfrak{L}_{3,0}$.} Similarly, this quantity comes from $\hat{H}_{3,0}$
\begin{equation}
\label{L30}
\begin{aligned}
& \mathfrak{L}_{3,0} \ =  \ \int_{-\infty}^0\mathrm{ds}\mathrm{Tr}\Big(\hat\rho_0[\hat{H}_{3,0},[f,\tilde{\hat{H}}_{3,0}(s)]]\Big)\\
&\approx  \   \frac{g^2n\pi}{V} \sum_{p_1,p_2,p_3,p_1',p_2',p_3'\ne 0}K^{3,0}_{1,2,3}K^{3,0}_{1',2',3'}\\
&\times\delta(\omega(p_1)+\omega(p_2)+\omega(p_3))\\
&\times \delta(p_1+p_2+p_3)\delta(p_1'+p_2'+p_3')\\
&\times \mathrm{Tr}\Big(\hat\rho_0\Big[[\hat{b}^\dagger_{p_3'}\hat{b}^\dagger_{p_2'}\hat{b}^\dagger_{p_1'},[\hat{b}^\dagger_p\hat{b}_p,\hat{b}_{p_1}\hat{b}_{p_2}\hat{b}_{p_3}]]\\
&+ [\hat{b}_{p_1'} \hat{b}_{p_2'}\hat{b}_{p_3'},[\hat{b}_p^\dagger \hat{b}_p,\hat{b}_{p_3}^\dagger \hat{b}_{p_2}^\dagger \hat{b}_{p_1}^\dagger]]\Big]\Big).
\end{aligned}
\end{equation}
 $K^{3,0}_{1',2',3'}$ has the same formulation with $K^{3,0}_{1,2,3}$, in which $p_1,p_2,p_3$  replaced by $p_1',p_2',p_3'$.

Since $\omega(p)>0$ for $p\ne 0$, the equation $\omega(p_1)+\omega(p_2)+\omega(p_3)=0$ does not have any solution. The quantity $\mathfrak{L}_{3,0}$ is then $0$.

{\it The form of $\mathfrak{L}_{2,2}$.} This quantity comes from $\hat{H}_{2,2}$
\begin{equation}
\label{L22}
\begin{aligned}
& \mathfrak{L}_{2,2} \ =  \ \int_{-\infty}^0\mathrm{ds}\mathrm{Tr}\Big(\hat\rho_0[\hat{H}_{2,2},[f,\tilde{\hat{H}}_{2,2}(s)]]\Big)\\
& \approx  \ \frac{g^2\pi}{4V^2}  \sum_{p_1,p_2,p_3,p_4,p_1',p_2',p_3',p_4'\ne 0}K^{2,2}_{1,2,3,4}K^{2,2}_{1',2',3',4'}\\
&\times\delta(\omega(p_1)+\omega(p_2)-\omega(p_3)-\omega(p_4))\\
& \times \delta(p_1+p_2-p_3-p_4)\delta(p_1'+p_2'-p_3'-p_4')\\
& \times \mathrm{Tr}\Big(\hat\rho_0[\hat{b}_{p_1'}^\dagger \hat{b}_{p_2'}^\dagger \hat{b}_{p_3'}\hat{b}_{p_4'}, [\hat{b}_{p}^\dagger \hat{b}_{p},\hat{b}_{p_1}^\dagger \hat{b}_{p_2}^\dagger \hat{b}_{p_3}\hat{b}_{p_4}]]\Big).
\end{aligned}
\end{equation}
 $K^{2,2}_{1',2',3',4'}$ has the same formulation with $K^{2,2}_{1,2,3,4}$, in which $p_1,p_2,p_3,p_4$ are replaced by $p_1',p_2',p_3',p_4'$.

{\it The form of $\mathfrak{L}_{3,1}$.} This quantity comes from $\hat{H}_{3,1}$
\begin{equation}
\label{L31}
\begin{aligned}
& \mathfrak{L}_{3,1} \ =  \ \int_{-\infty}^0\mathrm{ds}\mathrm{Tr}\Big(\hat\rho_0[\hat{H}_{3,1},[f,\tilde{\hat{H}}_{3,1}(s)]]\Big)\\
& \approx  \ \frac{g^2\pi}{4V^2}  \sum_{p_1,p_2,p_3,p_4,p_1',p_2',p_3',p_4'\ne 0}K^{3,1}_{1,2,3,4}K^{3,1}_{1',2',3',4'}\\
&\times \delta(\omega(p_1)-\omega(p_2)-\omega(p_3)-\omega(p_4))\\
&\times \delta(p_1-p_2-p_3-p_4)\delta(p_1'-p_2'-p_3'-p_4')\\
& \times \mathrm{Tr}\Big(\hat\rho_0\Big[[\hat{b}^\dagger_{p_4'}\hat{b}^\dagger_{p_3'}\hat{b}_{p_2'}^\dagger\hat{b}_{p_1'},[\hat{b}^\dagger_p\hat{b}_p,\hat{b}_{p_1}^\dagger \hat{b}_{p_2} \hat{b}_{p_3}\hat{b}_{p_4}]]\\
& + [\hat{b}^\dagger_{p_1'}\hat{b}_{p_2'}\hat{b}_{p_3'}\hat{b}_{p_4'},[\hat{b}^\dagger_p\hat{b}_p,\hat{b}_{p_4}^\dagger \hat{b}_{p_3}^\dagger \hat{b}_{p_2}^\dagger\hat{b}_{p_1}]]\Big]\Big).
\end{aligned}
\end{equation}
 $K^{3,1}_{1',2',3',4'}$ has the same formulation with $K^{3,1}_{1,2,3,4}$, in which $p_1,p_2,p_3,p_4$ are replaced by $p_1',p_2',p_3',p_4'$.

{\it The form of $\mathfrak{L}_{4,0}$.} This quantity comes from $\hat{H}_{4,0}$
\begin{equation}
\label{L40}
\begin{aligned}
&  \mathfrak{L}_{4,0} \ = \ \int_{-\infty}^0\mathrm{ds}\mathrm{Tr}\Big(\hat\rho_0[\hat{H}_{4,0},[f,\tilde{\hat{H}}_{4,0}(s)]]\Big)\\
&  \approx \ \frac{g^2\pi}{4V^2}  \sum_{p_1,p_2,p_3,p_4,p_1',p_2',p_3',p_4'\ne 0}K^{4,0}_{1,2,3,4}K^{4,0}_{1',2',3',4'}\\
&\times \delta(\omega(p_1)+\omega(p_2)+\omega(p_3)+\omega(p_4))\\
& \times \delta(p_1+p_2+p_3+p_4)\delta(p_1'+p_2'+p_3'+p_4')\\
& \times \mathrm{Tr}\Big(\hat\rho_0\Big[[\hat{b}^\dagger_{p_4'}\hat{b}^\dagger_{p_3'}\hat{b}^\dagger_{p_2'}\hat{b}^\dagger_{p_1'},[\hat{b}^\dagger_p\hat{b}_p,\hat{b}_{p_1} \hat{b}_{p_2} \hat{b}_{p_3}\hat{b}_{p_4}]]\\
&\indent\indent\indent + [\hat{b}_{p_1'}\hat{b}_{p_2'}\hat{b}_{p_3'}\hat{b}_{p_4'},[\hat{b}^\dagger_p\hat{b}_p,\hat{b}_{p_4}^\dagger \hat{b}^\dagger_{p_3} \hat{b}^\dagger_{p_2}\hat{b}^\dagger_{p_1}]] \Big]\Big).
\end{aligned}
\end{equation}
 $K^{4,0}_{1',2',3',4'}$ has the same formulation with $K^{4,0}_{1,2,3,4}$, in which $p_1,p_2,p_3,p_4$ are replaced by $p_1',p_2',p_3',p_4'$. Since $\omega(p)>0$ for $p\ne 0$, the equation $\omega(p_1)+\omega(p_2)+\omega(p_3)+\omega(p_4)=0$ does not have any solution. The quantity $\mathfrak{L}_{4,0}$ is indeed $0$. 

{\it The form of $\mathfrak{L}_{2,0}$.} This quantity comes from $\hat{H}_{2,0}$
\begin{equation}
\label{L20}
\begin{aligned}
& \mathfrak{L}_{2,0} \ =  \ \int_{-\infty}^0\mathrm{ds}\mathrm{Tr}\Big(\hat\rho_0[\hat{H}_{2,0},[f,\tilde{\hat{H}}_{2,0}(s)]]\Big)\\\
& \approx  \ \frac{g^2\pi}{4V^2}  \sum_{p_1,p_2,p_1',p_2'\ne 0}K^{2,0}_{1,2}K^{2,0}_{1',2'}\delta(\omega(p_1)+\omega(-p_1))\\
&\times \mathrm{Tr}\Big(\hat\rho_0\Big[[\hat{b}^\dagger_{p_1'}\hat{b}^\dagger_{-p_1'},[\hat{b}^\dagger_p\hat{b}_p,\hat{b}_{p_1} \hat{b}_{-p_1}]]\\
&+[\hat{b}_{p_1'}\hat{b}_{-p_1'},[\hat{b}^\dagger_p\hat{b}_p,\hat{b}^\dagger_{p_1} \hat{b}^\dagger_{-p_1}]]\Big]\Big).
\end{aligned}
\end{equation}
 $K^{2,0}_{1',2'}$ has the same formulation with $K^{2,0}_{1,2}$, in which $p_1,p_2$ are replaced by $p_1',p_2'$. Since $\omega(p)>0$ for $p\ne 0$, the equation $\omega(p_1)+\omega(-p_1)=0$ does not have any solution. The quantity $\mathfrak{L}_{2,0}$ is again $0$. 

{\it The form of $\mathfrak{L}_{1,1}$.} This quantity comes from $\hat{H}_{1,1}$
\begin{equation}
\label{L11}
\begin{aligned}
\mathfrak{L}_{1,1} \ = & \ \int_{-\infty}^0\mathrm{ds}\mathrm{Tr}\Big(\hat\rho_0[\hat{H}_{1,1},[f,\tilde{\hat{H}}_{1,1}(s)]]\Big)\\
\ \approx & \ \frac{g^2\pi}{4V^2}  \sum_{p_1,p_2,p_1',p_2'\ne 0}K^{1,1}_{1,2}K^{1,1}_{1',2'}\\
&\indent \times \int_{-\infty}^0\mathrm{ds}\mathrm{Tr}\Big(\hat\rho_0\Big[\hat{b}^\dagger_{p_1'}\hat{b}_{p_1'},[\hat{b}^\dagger_p\hat{b}_p,\hat{b}_{p_1}^\dagger \hat{b}_{p_1}]\Big]\Big).
\end{aligned}
\end{equation}
 $K^{1,1}_{1',2'}$ has the same formulation with $K^{1,1}_{1,2}$, in which $p_1,p_2$ are replaced by $p_1',p_2'$. This quantity is also $0$ due to the fact that 
$$\Big[\hat{b}^\dagger_{p_1'}\hat{b}_{p_1'},[\hat{b}^\dagger_p\hat{b}_p,\hat{b}_{p_1}^\dagger \hat{b}_{p_1}]\Big]=0.$$

We finally obtain the spatial homogeneous  equation
\begin{equation}
\label{LiouvilleFinal}
\begin{aligned}
& \partial_t f =   \frac{g^2n\pi}{V} \sum_{p_1,p_2,p_3,p_1',p_2',p_3'\ne 0}K^{1,2}_{1,2,3}K^{1,2}_{1',2',3'}\\
&\times \delta(p_1-p_2-p_3)\delta(p_1'-p_2'-p_3')\delta(\omega(p_1)-\omega(p_2)-\omega(p_3))\\
&\times\Big\langle \Big[[[\hat{b}^\dagger_{p_3'}\hat{b}^\dagger_{p_2'}\hat{b}_{p_1'},[\hat{b}^\dagger_p\hat{b}_p,\hat{b}^\dagger_{p_1}\hat{b}^\dagger_{p_2}\hat{b}_{p_3}]]\\
&+ [\hat{b}_{p_1'}^\dagger \hat{b}_{p_2'}\hat{b}_{p_3'},[\hat{b}_p^\dagger \hat{b}_p,\hat{b}_{p_3}^\dagger \hat{b}_{p_2}^\dagger \hat{b}_{p_1}]]\Big]\Big\rangle\\
 & + \ \frac{g^2\pi}{4V^2}  \sum_{p_1,p_2,p_3,p_4,p_1',p_2',p_3',p_4'\ne 0}K^{2,2}_{1,2,3,4}K^{2,2}_{1',2',3',4'}\\
& \times \delta(p_1+p_2-p_3-p_4)\delta(p_1'+p_2'-p_3'-p_4')\\
&\times\delta(\omega(p_1)+\omega(p_2)-\omega(p_3)-\omega(p_4))\\
&\times\left\langle [\hat{b}_{p_1'}^\dagger \hat{b}_{p_2'}^\dagger \hat{b}_{p_3'}\hat{b}_{p_4'}, [\hat{b}_{p}^\dagger \hat{b}_{p},\hat{b}_{p_1}^\dagger \hat{b}_{p_2}^\dagger \hat{b}_{p_3}\hat{b}_{p_4}]]\right\rangle\\
&+ \frac{g^2\pi}{4V^2}  \sum_{p_1,p_2,p_3,p_4,p_1',p_2',p_3',p_4'\ne 0}K^{3,1}_{1,2,3,4}K^{3,1}_{1',2',3',4'}\\
&\times\delta(\omega(p_1)-\omega(p_2)-\omega(p_3)-\omega(p_4))\\
& \times \delta(p_1-p_2-p_3-p_4)\delta(p_1'-p_2'-p_3'-p_4')\\
&\times \left\langle\Big[[\hat{b}^\dagger_{p_4'}\hat{b}^\dagger_{p_3'}\hat{b}^\dagger_{p_2'}\hat{b}_{p_1'},[\hat{b}^\dagger_p\hat{b}_p,\hat{b}_{p_1}^\dagger \hat{b}_{p_2}\hat{b}_{p_3}\hat{b}_{p_4}]]\right.\\
&\indent\indent\indent\left.+ [\hat{b}^\dagger_{p_1'}\hat{b}_{p_2'}\hat{b}_{p_3'}\hat{b}_{p_4'},[\hat{b}^\dagger_p\hat{b}_p,\hat{b}_{p_4}^\dagger \hat{b}^\dagger_{p_3}\hat{b}^\dagger_{p_2}\hat{b}_{p_1}]]\Big]\right\rangle.
\end{aligned}
\end{equation}
\section{The kinetic equation}
The left hand side of \eqref{LiouvilleFinal} is split into  three terms, each contains  special types  of commutators. The first term includes commutators of the types $[\hat{b}^\dagger \hat{b}^\dagger \hat{b},[\hat{b}^\dagger \hat{b},\hat{b}^\dagger \hat{b} \hat{b}]]$ and $[\hat{b}^\dagger \hat{b}\hat{b} ,[\hat{b}^\dagger \hat{b},\hat{b}^\dagger \hat{b}^\dagger \hat{b}]]$. The computations in Appendix C show that this term is indeed the $C_{12}$ collision operator. The second term has commutators of the type $[\hat{b}^\dagger \hat{b}^\dagger \hat{b}\hat{b},[\hat{b}^\dagger \hat{b},\hat{b}^\dagger \hat{b}^\dagger \hat{b}\hat{b}]]$. This term can be shown to be  the collision operator $C_{22}$. The explicit computations of this collision operator is postponed to Appendix D. The $C_{31}$ collision operator comes from the last term, which involves commutators of the types $[\hat{b}^\dagger \hat{b}^\dagger \hat{b}^\dagger\hat{b},[\hat{b}^\dagger \hat{b},\hat{b}^\dagger \hat{b} \hat{b}\hat{b}]]$ and $[\hat{b}^\dagger \hat{b}\hat{b}\hat{b},[\hat{b}^\dagger\hat{b},\hat{b}^\dagger \hat{b}^\dagger \hat{b}^\dagger \hat{b}]]$. These computations are given in detail in Appendix E. In conclusion, by computing explicitly the commutators on the right hand side of \eqref{LiouvilleFinal}, we finally arrive at the kinetic equation  \eqref{KineticFinal}.
We emphasize that the density of the thermal cloud $f(t,p)$  is defined only for $p\ne 0$  due to  the fact that the condensate has been factored out  in the Bogoliubov diagonalization  \eqref{hata0}.
\\{\bf Acknowledgements.} 
 We would like to thank Prof. Linda Reichl for carefully reading the manuscript and Prof. Jay Robert Dorfman for  fruitful discussions on the topic.  M.-B. Tran is partially supported by NSF Grant DMS-1814149, NSF Grant DMS-1854453, SMU URC Grant 2020. 
 \section{Appendices}
 \subsection{Appendix A}
Let us expand $\delta (p_1+p_2-p_3)\hat{a}_{p_1}^\dagger \hat{a}_{p_2}^\dagger \hat{a}_{p_3}$ in terms of $\hat{b}_{p_1}^\dagger $, $\hat{b}_{p_1}$, $\hat{b}_{p_2}^\dagger $, $\hat{b}_{p_2}$, $\hat{b}_{p_3}^\dagger $, $\hat{b}_{p_3}$ 
\begin{equation}\label{H2Eq1}
\begin{aligned}
&\delta (p_1+p_2-p_3) \hat{a}_{p_1}^\dagger \hat{a}_{p_2}^\dagger \hat{a}_{p_3}  
\\
\ = & \ \delta (p_1+p_2-p_3) [u_{p_1} \hat{b}_{p_1}^\dagger - v_{p_1}\hat{b}_{-p_1}][u_{p_2} \hat{b}_{p_2}^\dagger - v_{p_2}\hat{b}_{-p_2}]\\
&\ \times [u_{p_3} \hat{b}_{p_3} - v_{p_3}\hat{b}_{-p_3}^\dagger]\\
 \ = & \ \delta (p_1+p_2-p_3) \Big[u_{p_1}u_{p_2}u_{p_3}\hat{b}_{p_1}^\dagger \hat{b}_{p_2}^\dagger \hat{b}_{p_3} -v_{p_1}u_{p_2}u_{p_3}\hat{b}_{-p_1} \hat{b}_{p_2}^\dagger \hat{b}_{p_3} \\
 & \  -u_{p_1}v_{p_2}u_{p_3}\hat{b}_{p_1}^\dagger \hat{b}_{-p_2} \hat{b}_{p_3}   + v_{p_1}v_{p_2}u_{p_3}\hat{b}_{-p_1} \hat{b}_{-p_2} \hat{b}_{p_3} \\
&\ \indent  - u_{p_1}u_{p_2}v_{p_3}\hat{b}_{p_1}^\dagger \hat{b}_{p_2}^\dagger \hat{b}_{-p_3}^\dagger   +  v_{p_1}u_{p_2}v_{p_3}\hat{b}_{-p_1} \hat{b}_{p_2}^\dagger \hat{b}_{-p_3}^\dagger  \\
&\ \indent  + u_{p_1}v_{p_2}v_{p_3}\hat{b}_{p_1}^\dagger \hat{b}_{-p_2}\hat{b}_{-p_3}^\dagger   -  v_{p_1}v_{p_2}v_{p_3}\hat{b}_{-p_1} \hat{b}_{-p_2}\hat{b}_{-p_3}^\dagger\Big].
\end{aligned}
\end{equation}

Similarly, we expand $\delta(p_1-p_2-p_3)\hat{a}_{p_1}^\dagger \hat{a}_{p_2} \hat{a}_{p_3}$ in terms of $\hat{b}_{p_1}^\dagger $, $\hat{b}_{p_1}$, $\hat{b}_{p_2}^\dagger $, $\hat{b}_{p_2}$, $\hat{b}_{p_3}^\dagger $, $\hat{b}_{p_3}$ 
\begin{equation}\label{H2Eq2}
\begin{aligned}
&\delta (p_1-p_2-p_3) \hat{a}_{p_1}^\dagger \hat{a}_{p_2} \hat{a}_{p_3}  
\\
\ = & \delta (p_1-p_2-p_3) [u_{p_1} \hat{b}_{p_1}^\dagger - v_{p_1}\hat{b}_{-p_1}][u_{p_2} \hat{b}_{p_2} - v_{p_2}\hat{b}_{-p_2}^\dagger]\\
& \times [u_{p_3} \hat{b}_{p_3} - v_{p_3}\hat{b}_{-p_3}^\dagger]\\
 \ = &\delta (p_1-p_2-p_3) \Big[u_{p_1}u_{p_2}u_{p_3}\hat{b}_{p_1}^\dagger \hat{b}_{p_2}\hat{b}_{p_3}   -  v_{p_1}u_{p_2}u_{p_3}\hat{b}_{-p_1} \hat{b}_{p_2}\hat{b}_{p_3} \\
 & \  -u_{p_1}v_{p_2}u_{p_3}\hat{b}_{p_1}^\dagger \hat{b}_{-p_2}^\dagger \hat{b}_{p_3}   + v_{p_1}v_{p_2}u_{p_3}\hat{b}_{-p_1} \hat{b}_{-p_2}^\dagger \hat{b}_{p_3} \\
&\   - u_{p_1}u_{p_2}v_{p_3}\hat{b}_{p_1}^\dagger \hat{b}_{p_2}\hat{b}_{-p_3}^\dagger   +  v_{p_1}u_{p_2}v_{p_3}\hat{b}_{-p_1} \hat{b}_{p_2} \hat{b}_{-p_3}^\dagger  \\
&\    + u_{p_1}v_{p_2}v_{p_3}\hat{b}_{p_1}^\dagger \hat{b}_{-p_2}^\dagger \hat{b}_{-p_3}^\dagger   -  v_{p_1}v_{p_2}v_{p_3}\hat{b}_{-p_1} \hat{b}_{-p_2}^\dagger \hat{b}_{-p_3}^\dagger\Big].
\end{aligned}
\end{equation}
We perform the following change of variables, for the terms in \eqref{H2Eq1}, taking into account the fact that $p_1,p_2,p_3\ne 0$
\begin{widetext}
\begin{equation}\label{H2Eq3}
\begin{aligned}
\delta (p_1+p_2-p_3) u_{p_1}u_{p_2}u_{p_3}\hat{b}_{p_1}^\dagger \hat{b}_{p_2}^\dagger \hat{b}_{p_3}
  \to &  \delta (p_1-p_2-p_3) u_{p_1}u_{p_2}u_{p_3}\hat{b}_{p_3}^\dagger \hat{b}_{p_2}^\dagger \hat{b}_{p_1},\\
-  \delta (p_1+p_2-p_3) v_{p_1}u_{p_2}u_{p_3}\hat{b}_{-p_1} \hat{b}_{p_2}^\dagger \hat{b}_{p_3}
 \to & -   \delta (p_1-p_2-p_3) u_{p_1}v_{p_2}u_{p_3}\hat{b}_{p_2}\hat{b}_{p_1}^\dagger \hat{b}_{p_3}
 = -  \delta (p_1-p_2-p_3) u_{p_1}v_{p_2}u_{p_3}\hat{b}_{p_1}^\dagger \hat{b}_{p_2} \hat{b}_{p_3},\\
- \delta (p_1+p_2-p_3) u_{p_1}v_{p_2}u_{p_3}\hat{b}_{p_1}^\dagger  \hat{b}_{-p_2}\hat{b}_{p_3} \to & - \ \delta (p_1-p_2-p_3) u_{p_1}u_{p_2}v_{p_3}\hat{b}_{p_1}^\dagger \hat{b}_{p_2}\hat{b}_{p_3},\\
  \delta (p_1+p_2-p_3) v_{p_1}v_{p_2}u_{p_3}\hat{b}_{-p_1} \hat{b}_{-p_2} \hat{b}_{p_3} 
 \to &   \delta (p_1-p_2-p_3) u_{p_1}v_{p_2}v_{p_3}\hat{b}_{-p_1} \hat{b}_{p_2}\hat{b}_{p_3},\\
\delta (p_1+p_2-p_3) v_{p_1}u_{p_2}v_{p_3}\hat{b}_{-p_1}  \hat{b}_{p_2}^\dagger \hat{b}_{-p_3}^\dagger \to &  \delta (p_1-p_2-p_3) v_{p_1} u_{p_2}v_{p_3}\hat{b}_{p_1}  \hat{b}_{p_2}^\dagger \hat{b}_{p_3}^\dagger
  =  \delta (p_1-p_2-p_3) v_{p_1}u_{p_2}v_{p_3}   \hat{b}_{p_3}^\dagger \hat{b}_{p_2}^\dagger \hat{b}_{p_1},\\
  \delta (p_1+p_2-p_3)  u_{p_1}v_{p_2}v_{p_3}\hat{b}_{p_1}^\dagger \hat{b}_{-p_2}\hat{b}_{-p_3}^\dagger \to &\ \delta (p_1-p_2-p_3)  v_{p_1}u_{p_2}v_{p_3}\hat{b}_{p_2}^\dagger \hat{b}_{p_1}\hat{b}_{p_3}^\dagger=   \delta (p_1-p_2-p_3)  v_{p_1}v_{p_2}u_{p_3}\hat{b}_{p_3}^\dagger \hat{b}_{p_2}^\dagger \hat{b}_{p_1},\\
 - \delta (p_1+p_2-p_3) v_{p_1}v_{p_2}v_{p_3}\hat{b}_{-p_1} \hat{b}_{-p_2}\hat{b}_{-p_3}^\dagger 
\to & -\delta (p_1-p_2-p_3) v_{p_1}v_{p_2}v_{p_3} \hat{b}_{p_2} \hat{b}_{p_3}\hat{b}_{p_1}^\dagger= -  \delta (p_1-p_2-p_3) v_{p_1}v_{p_2}v_{p_3} \hat{b}_{p_1}^\dagger \hat{b}_{p_2} \hat{b}_{p_3},\\
-  \delta (p_1+p_2-p_3) u_{p_1}u_{p_2}v_{p_3}\hat{b}_{p_1}^\dagger \hat{b}_{p_2}^\dagger \hat{b}_{-p_3}^\dagger  \to & - \ \delta (p_1-p_2-p_3) v_{p_1}u_{p_2}u_{p_3} \hat{b}_{p_3}^\dagger \hat{b}_{p_2}^\dagger \hat{b}_{-p_1}^\dagger.
\end{aligned}
\end{equation}
\end{widetext}
Notice that in the above computations, we used identities like $\delta(p_1-p_2-p_3) \hat{b}_{p_2}\hat{b}_{p_1}^\dagger \hat{b}_{p_3} = \delta(p_1-p_2-p_3) \hat{b}_{p_1}^\dagger \hat{b}_{p_2} \hat{b}_{p_3} + \delta(p_1-p_2-p_3) \delta(p_1-p_2)\hat{b}_{p_3} = \delta(p_1-p_2-p_3) \hat{b}_{p_1}^\dagger \hat{b}_{p_2} \hat{b}_{p_3}$ due to the fact that all $p_1,p_2,p_3$ are non-zero. 
Similar computations can also be carried out for the terms in \eqref{H2Eq2}
\begin{widetext}
\begin{equation}\label{H2Eq4}
\begin{aligned}
-  \delta (p_1-p_2-p_3) u_{p_1}v_{p_2}u_{p_3}\hat{b}_{p_1} ^\dagger \hat{b}_{-p_2}^\dagger \hat{b}_{p_3}
  \to  - & \delta (p_1-p_2-p_3) u_{p_1}v_{p_2}u_{p_3}\hat{b}_{p_3}^\dagger \hat{b}_{p_2}^\dagger \hat{b}_{p_1},\\
 \delta (p_1-p_2-p_3) v_{p_1}v_{p_2}u_{p_3}\hat{b}_{-p_1} \hat{b}_{-p_2}^\dagger \hat{b}_{p_3}
 \to  &\  \delta (p_1-p_2-p_3) v_{p_1}v_{p_2}u_{p_3}\hat{b}_{p_2}\hat{b}_{p_1}^\dagger \hat{b}_{p_3}
 =   \delta (p_1-p_2-p_3) v_{p_1}v_{p_2}u_{p_3}\hat{b}_{p_1}^\dagger \hat{b}_{p_2} \hat{b}_{p_3},\\
- \delta (p_1-p_2-p_3) u_{p_1}u_{p_2}v_{p_3}\hat{b}_{p_1}^\dagger  \hat{b}_{p_2}\hat{b}_{-p_3}^\dagger 
\to   - & \delta (p_1-p_2-p_3)  u_{p_1}u_{p_2}v_{p_3}\hat{b}_{p_2}^\dagger  \hat{b}_{p_1}\hat{b}_{p_3}^\dagger
=   -  \delta (p_1-p_2-p_3) u_{p_1}u_{p_2}v_{p_3}\hat{b}_{p_3}^\dagger \hat{b}_{p_2}^\dagger  \hat{b}_{p_1},\\
 \delta (p_1-p_2-p_3) v_{p_1}u_{p_2}v_{p_3}\hat{b}_{-p_1}  \hat{b}_{p_2}\hat{b}_{-p_3}^\dagger
 \to  &  \delta (p_1-p_2-p_3) v_{p_1}u_{p_2}v_{p_3}\hat{b}_{p_2}  \hat{b}_{p_3} \hat{b}_{p_1}^\dagger
  =  \delta (p_1-p_2-p_3) v_{p_1}u_{p_2}v_{p_3}\hat{b}_{p_1}^\dagger \hat{b}_{p_2}  \hat{b}_{p_3},\\
 \delta (p_1-p_2-p_3)  u_{p_1}v_{p_2}v_{p_3}\hat{b}_{p_1}^\dagger \hat{b}_{-p_2}^\dagger \hat{b}_{-p_3}^\dagger
  \to  &\  \delta (p_1-p_2-p_3)  u_{p_1}v_{p_2}v_{p_3}\hat{b}_{p_3}^\dagger \hat{b}_{p_2}^\dagger \hat{b}_{-p_1}^\dagger,\\
 -  \delta (p_1-p_2-p_3) v_{p_1}v_{p_2}v_{p_3}\hat{b}_{-p_1} \hat{b}_{-p_2}^\dagger \hat{b}_{-p_3}^\dagger 
  \to & -  \delta (p_1-p_2-p_3) v_{p_1}v_{p_2}v_{p_3} \hat{b}_{p_1} \hat{b}_{p_2}^\dagger \hat{b}_{p_3}^\dagger\ = - \delta (p_1-p_2-p_3) v_{p_1}v_{p_2}v_{p_3} \hat{b}_{p_3}^\dagger \hat{b}_{p_2}^\dagger \hat{b}_{p_1}.
\end{aligned}
\end{equation}
\end{widetext}

We finally do the change of variables
\begin{widetext}
\begin{equation}\label{H2Eq4a}
\begin{aligned}
& \delta(p_1-p_2-p_3)\Big[(u_{p_1}v_{p_2}v_{p_3}-v_{p_1}u_{p_2}u_{p_3})(\hat{b}_{p_3}^\dagger \hat{b}_{p_2}^\dagger \hat{b}_{-p_1}^\dagger + \hat{b}_{-p_1} \hat{b}_{p_2}\hat{b}_{p_3})\Big]\\
\ = & \ \delta(p_1+p_2+p_3)\Big[(u_{p_1}v_{p_2}v_{p_3}-v_{p_1}u_{p_2}u_{p_3})(\hat{b}_{p_3}^\dagger \hat{b}_{p_2}^\dagger \hat{b}_{p_1}^\dagger + \hat{b}_{p_1} \hat{b}_{p_2}\hat{b}_{p_3})\Big].
\end{aligned}
\end{equation}
\end{widetext}

Putting together all of the identities in \eqref{H2Eq1}-\eqref{H2Eq4a} yields the new form \eqref{H2Eq5} of $\hat{H}_2$.
 \subsection{Appendix B}
 Let us now expand $\delta (p_1+p_2-p_3-p_4)\hat{a}_{p_1}^\dagger \hat{a}_{p_2}^\dagger \hat{a}_{p_3}\hat{a}_{p_4}$ in terms of $\hat{b}_{p_1}^\dagger $, $\hat{b}_{p_1}$, $\hat{b}_{p_2}^\dagger $, $\hat{b}_{p_2}$, $\hat{b}_{p_3}^\dagger $, $\hat{b}_{p_3}$, $\hat{b}_{p_4}^\dagger $, $\hat{b}_{p_4}$
\begin{widetext}
\begin{equation}\label{H3Eq1}
\begin{aligned}
&\delta (p_1+p_2-p_3-p_4) \hat{a}_{p_1}^\dagger \hat{a}_{p_2}^\dagger \hat{a}_{p_3}\hat{a}_{p_4}  
\\
\ = & \ \delta (p_1+p_2-p_3-p_4) [u_{p_1} \hat{b}_{p_1}^\dagger - v_{p_1}\hat{b}_{-p_1}][u_{p_2} \hat{b}_{p_2}^\dagger - v_{p_2}\hat{b}_{-p_2}][u_{p_3} \hat{b}_{p_3} - v_{p_3}\hat{b}_{-p_3}^\dagger][u_{p_4} \hat{b}_{p_4} - v_{p_4}\hat{b}_{-p_4}^\dagger]\\
 \ = & \ \delta (p_1+p_2-p_3-p_4) \Big[u_{p_1}u_{p_2}u_{p_3}u_{p_4}\hat{b}_{p_1}^\dagger \hat{b}_{p_2}^\dagger \hat{b}_{p_3}\hat{b}_{p_4} -  v_{p_1}u_{p_2}u_{p_3}u_{p_4}\hat{b}_{-p_1} \hat{b}_{p_2}^\dagger \hat{b}_{p_3}\hat{b}_{p_4} \\
 & -  u_{p_1}v_{p_2}u_{p_3}u_{p_4}\hat{b}_{p_1}^\dagger \hat{b}_{-p_2} \hat{b}_{p_3}\hat{b}_{p_4} +  v_{p_1}v_{p_2}u_{p_3}u_{p_4}\hat{b}_{-p_1} \hat{b}_{-p_2} \hat{b}_{p_3}\hat{b}_{p_4} -  u_{p_1}u_{p_2}v_{p_3}u_{p_4}\hat{b}_{p_1}^\dagger \hat{b}_{p_2}^\dagger \hat{b}_{-p_3}^\dagger \hat{b}_{p_4} \\
 &  +  v_{p_1}u_{p_2}v_{p_3}u_{p_4}\hat{b}_{-p_1} \hat{b}_{p_2}^\dagger \hat{b}_{-p_3}^\dagger \hat{b}_{p_4} +  u_{p_1}v_{p_2}v_{p_3}u_{p_4}\hat{b}_{p_1}^\dagger \hat{b}_{-p_2} \hat{b}_{-p_3}^\dagger \hat{b}_{p_4}-  v_{p_1}v_{p_2}v_{p_3}u_{p_4}\hat{b}_{-p_1} \hat{b}_{-p_2}\hat{b}_{-p_3}^\dagger \hat{b}_{p_4} \\
   &\ - \ u_{p_1}u_{p_2}u_{p_3}v_{p_4}\hat{b}_{p_1}^\dagger \hat{b}_{p_2}^\dagger \hat{b}_{p_3}\hat{b}_{-p_4}^\dagger + v_{p_1}u_{p_2}u_{p_3}v_{p_4}\hat{b}_{-p_1} \hat{b}_{p_2}^\dagger \hat{b}_{p_3} \hat{b}_{-p_4}^\dagger  +  u_{p_1}v_{p_2}u_{p_3}v_{p_4}\hat{b}_{p_1}^\dagger \hat{b}_{-p_2} \hat{b}_{p_3}\hat{b}_{-p_4}^\dagger  \\
 &  -  v_{p_1}v_{p_2}u_{p_3}v_{p_4}\hat{b}_{-p_1} \hat{b}_{-p_2} \hat{b}_{p_3} \hat{b}_{-p_4}^\dagger  + u_{p_1}u_{p_2}v_{p_3}v_{p_4}\hat{b}_{p_1}^\dagger \hat{b}_{p_2}^\dagger \hat{b}_{-p_3}^\dagger \hat{b}_{-p_4}^\dagger  -  v_{p_1}u_{p_2}v_{p_3}v_{p_4}\hat{b}_{-p_1} \hat{b}_{p_2}^\dagger \hat{b}_{-p_3}^\dagger \hat{b}_{-p_4}^\dagger \\
 &-  u_{p_1}v_{p_2}v_{p_3}v_{p_4}\hat{b}_{p_1}^\dagger \hat{b}_{-p_2} \hat{b}_{-p_3}^\dagger \hat{b}_{-p_4}^\dagger +  v_{p_1}v_{p_2}v_{p_3}v_{p_4}\hat{b}_{-p_1} \hat{b}_{-p_2}\hat{b}_{-p_3}^\dagger \hat{b}_{-p_4}^\dagger \Big].
\end{aligned}
\end{equation}
\end{widetext}
Similarly as for $\hat{H}_2$, we perform several changes of variables, in combination with evaluating the commutators, to obtain
\begin{widetext}
\begin{equation}
\begin{aligned}\label{H3Eq2}
 & \delta(p_1+p_2-p_3-p_4)u_{p_1}v_{p_2}v_{p_3}u_{p_4}\hat{b}_{p_1}^\dagger \hat{b}_{-p_2} \hat{b}_{-p_3}^\dagger \hat{b}_{p_4} 
    \  =\ \delta(p_1+p_2-p_3-p_4)u_{p_1}v_{p_2}v_{p_3}u_{p_4}\Big[\hat{b}_{p_1}^\dagger  \hat{b}_{-p_3}^\dagger \hat{b}_{-p_2} \hat{b}_{p_4}\\
   &\indent\indent\indent\indent\indent\indent\indent\indent\indent\indent\indent\indent\indent\indent\indent\indent\indent\indent\indent\indent\indent\indent\indent+ \ \delta(p_2-p_3)\hat{b}_{p_1}^\dagger \hat{b}_{p_4}\Big]\\
  \indent   \to &\ \delta(p_1+p_2-p_3-p_4)u_{p_1}v_{p_2}v_{p_3}u_{p_4}\hat{b}_{p_1}^\dagger  \hat{b}_{p_2}^\dagger \hat{b}_{p_3} \hat{b}_{p_4}+ \ u_{p_1}^2 v_{p_2}^2 \hat{b}_{p_1}^\dagger \hat{b}_{p_1},
\\
       ~~ & ~~ \\ 
&\delta(p_1+p_2-p_3-p_4) u_{p_1}v_{p_2}u_{p_3}v_{p_4}\hat{b}_{p_1}^\dagger \hat{b}_{-p_2}\hat{b}_{p_3}\hat{b}_{-p_4}^\dagger\ 
 = \ \delta(p_1+p_2-p_3-p_4)u_{p_1}v_{p_2}u_{p_3}v_{p_4}\Big[\hat{b}_{p_1}^\dagger \hat{b}_{-p_4}^\dagger \hat{b}_{-p_2}\hat{b}_{p_3}\\
 &\indent\indent\indent\indent\indent\indent\indent\indent\indent\indent\indent\indent\indent\indent\indent\indent\indent\indent\indent\indent\indent\indent\indent\indent + \delta(p_2-p_4)\hat{b}_{p_1}^\dagger \hat{b}_{p_3} +\delta(p_3+p_4)\hat{b}_{p_1}^\dagger \hat{b}_{-p_2}\Big]\\
   \to & \ \delta(p_1+p_2-p_3-p_4) u_{p_1}v_{p_2}u_{p_3}v_{p_4}\hat{b}_{p_1}^\dagger  \hat{b}_{p_2}^\dagger \hat{b}_{p_3} \hat{b}_{p_4} +  (u_{p_1}^2 v_{p_2}^2 + u_{p_1}v_{p_1}u_{p_2}v_{p_2})\hat{b}_{p_1}^\dagger \hat{b}_{p_1},\\
          ~~ & ~~ \\ 
 &\delta(p_1+p_2-p_3-p_4) v_{p_1}u_{p_2}v_{p_3}u_{p_4}\hat{b}_{-p_1} \hat{b}_{p_2}^\dagger \hat{b}_{-p_3}^\dagger \hat{b}_{p_4}\
  = \ \delta(p_1+p_2-p_3-p_4)v_{p_1}u_{p_2}v_{p_3}u_{p_4}\\
   &\indent\indent\indent\indent\indent\indent\indent\indent\indent\indent\indent\indent\indent\indent\times \Big[\hat{b}_{p_2}^\dagger \hat{b}_{-p_3}^\dagger \hat{b}_{-p_1}\hat{b}_{p_4} + \delta(p_2+p_1)\hat{b}_{-p_3}^\dagger \hat{b}_{p_4}\ +\ \delta(p_1-p_3)\hat{b}_{p_2}^\dagger \hat{b}_{p_4}\Big]\\
    \to &\ \delta(p_1+p_2-p_3-p_4) v_{p_1}u_{p_2}v_{p_3}u_{p_4}\hat{b}_{p_1}^\dagger  \hat{b}_{p_2}^\dagger \hat{b}_{p_3} \hat{b}_{p_4} \
 + \ (u_{p_1}^2 v_{p_2}^2 + u_{p_1}v_{p_1}u_{p_2}v_{p_2})\hat{b}_{p_1}^\dagger \hat{b}_{p_1},\\
           ~~ & ~~ \\ 
 &\delta(p_1+p_2-p_3-p_4) v_{p_1}u_{p_2}u_{p_3}v_{p_4}\hat{b}_{-p_1} \hat{b}_{p_2}^\dagger \hat{b}_{p_3}\hat{b}_{-p_4}^\dagger \
  = \ \delta(p_1+p_2-p_3-p_4)v_{p_1}u_{p_2}u_{p_3}v_{p_4}\\
  & \times \Big[\hat{b}_{p_2}^\dagger \hat{b}_{-p_4}^\dagger \hat{b}_{-p_1}\hat{b}_{p_3} + \delta(p_1+p_2)\hat{b}_{p_3} \hat{b}_{-p_4}^\dagger   \ +\ \delta(p_3+p_4)\hat{b}_{p_2}^\dagger \hat{b}_{-p_1} +\delta(p_1-p_4)\hat{b}_{p_2}^\dagger \hat{b}_{p_3}\Big]\\
   \to  & \ \delta(p_1+p_2-p_3-p_4) v_{p_1}u_{p_2}u_{p_3}v_{p_4} \hat{b}_{p_1}^\dagger  \hat{b}_{p_2}^\dagger \hat{b}_{p_3} \hat{b}_{p_4}
   +  (u_{p_1}^2 v_{p_2}^2 + u_{p_1}v_{p_1}u_{p_2}v_{p_2})\hat{b}_{p_1}^\dagger \hat{b}_{p_1} +u_{p_1}v_{p_1}u_{p_2}v_{p_2}\hat{b}_{p_1} \hat{b}_{p_1}^\dagger,\\
              ~~ & ~~ \\ 
 &\delta(p_1+p_2-p_3-p_4) v_{p_1}v_{p_2}v_{p_3}v_{p_4}\hat{b}_{-p_1} \hat{b}_{-p_2} \hat{b}_{-p_3}^\dagger \hat{b}_{-p_4}^\dagger \
  \to\  \delta(p_1+p_2-p_3-p_4) v_{p_1}v_{p_2}v_{p_3}v_{p_4}\hat{b}_{p_1} \hat{b}_{p_2} \hat{b}_{p_3}^\dagger \hat{b}_{p_4}^\dagger \\
 = & \   \delta(p_1+p_2-p_3-p_4)v_{p_1}v_{p_2}v_{p_3}v_{p_4}\Big[\hat{b}_{p_3}^\dagger \hat{b}_{p_4}^\dagger \hat{b}_{p_1}\hat{b}_{p_2} + \delta(p_2-p_3)\hat{b}_{p_1} \hat{b}_{p_4}^\dagger + \delta(p_2-p_4)\hat{b}_{p_1} \hat{b}_{p_3}^\dagger \ +\delta(p_1-p_3)\hat{b}_{p_4}^\dagger \hat{b}_{p_2}\\
 &\indent\indent\indent\indent\indent\indent\indent\indent\indent\indent\indent\indent\indent\indent +\delta(p_1-p_4)\hat{b}_{p_3}^\dagger \hat{b}_{p_2}\Big]\\
    \to &\ \delta(p_1+p_2-p_3-p_4) v_{p_1}v_{p_2}v_{p_3}v_{p_4} \hat{b}_{p_1}^\dagger  \hat{b}_{p_2}^\dagger \hat{b}_{p_3} \hat{b}_{p_4}  +  2v_{p_1}^2 v_{p_2}^2 \hat{b}_{p_1}^\dagger \hat{b}_{p_1} +  2v_{p_1}^2 v_{p_2}^2 \hat{b}_{p_1} \hat{b}_{p_1}^\dagger,
\end{aligned}
\end{equation}
\end{widetext}
as well as
\begin{widetext}
\begin{equation}
\begin{aligned}\label{H3Eq3}
& \delta(p_1+p_2-p_3-p_4) v_{p_1}u_{p_2}u_{p_3}u_{p_4}\hat{b}_{-p_1}\hat{b}_{p_2}^\dagger \hat{b}_{p_3}\hat{b}_{p_4}  \to \delta(p_1-p_2-p_3-p_4) u_{p_1}v_{p_2}u_{p_3}u_{p_4}\hat{b}_{p_2}\hat{b}_{p_1}^\dagger \hat{b}_{p_3}\hat{b}_{p_4}  \\
= & \ \delta(p_1-p_2-p_3-p_4) u_{p_1}v_{p_2}u_{p_3}u_{p_4} \Big[\hat{b}_{p_1}^\dagger \hat{b}_{p_2} \hat{b}_{p_3}\hat{b}_{p_4} + \delta(p_1-p_2)\hat{b}_{p_3}\hat{b}_{p_4} \Big]\\
\to & \ \delta(p_1-p_2-p_3-p_4) u_{p_1}u_{p_2}v_{p_3}u_{p_4}\hat{b}_{p_1}^\dagger \hat{b}_{p_2} \hat{b}_{p_3}\hat{b}_{p_4} + u_{p_1}^2 u_{p_2}v_{p_2}\hat{b}_{p_1}\hat{b}_{-p_1},\\
              ~~ & ~~ \\ 
& \delta(p_1+p_2-p_3-p_4) u_{p_1}v_{p_2}u_{p_3}u_{p_4}\hat{b}_{p_1}^\dagger \hat{b}_{-p_2} \hat{b}_{p_3}\hat{b}_{p_4}\\
\to& \ \delta(p_1-p_2-p_3-p_4) u_{p_1}u_{p_2}v_{p_3}u_{p_4}\hat{b}_{p_1}^\dagger \hat{b}_{p_2}\hat{b}_{p_3}\hat{b}_{p_4},  \\
              ~~ &~~ \\ 
& \delta(p_1+p_2-p_3-p_4) u_{p_1}u_{p_2}v_{p_3}u_{p_4}\hat{b}_{p_1}^\dagger \hat{b}_{p_2}^\dagger  \hat{b}_{-p_3}^\dagger  \hat{b}_{p_4}
\to\delta(p_1-p_2-p_3-p_4) u_{p_1}u_{p_2}v_{p_3}u_{p_4}\hat{b}_{p_4}^\dagger \hat{b}_{p_3}^\dagger \hat{b}_{p_2}^\dagger \hat{b}_{p_1} , \\
                ~~ & ~~ \\ 
& \delta(p_1+p_2-p_3-p_4) v_{p_1}v_{p_2}v_{p_3}u_{p_4}\hat{b}_{-p_1} \hat{b}_{-p_2}  \hat{b}_{-p_3}^\dagger  \hat{b}_{p_4} 
\to  \delta(p_1-p_2-p_3-p_4) v_{p_1}v_{p_2}v_{p_3}u_{p_4} \hat{b}_{p_2} \hat{b}_{p_3} \hat{b}_{p_1} ^\dagger \hat{b}_{p_4} \\
 \ =& \ \delta(p_1-p_2-p_3-p_4) v_{p_1}v_{p_2}v_{p_3}u_{p_4}\Big[\hat{b}_{p_1} ^\dagger \hat{b}_{p_2} \hat{b}_{p_3}  \hat{b}_{p_4} + \delta{(p_1-p_3)}\hat{b}_{p_2}\hat{b}_{p_4} + \delta(p_1-p_2)\hat{b}_{p_3}\hat{b}_{p_4} \Big] \\
 \to & \ \delta(p_1-p_2-p_3-p_4) v_{p_1}v_{p_2}u_{p_3}v_{p_4} \hat{b}_{p_1} ^\dagger \hat{b}_{p_2} \hat{b}_{p_3}  \hat{b}_{p_4}\ +\ 2\hat{b}_{p_1}\hat{b}_{-p_1} v_{p_1}u_{p_1}v_{p_2}^2,
\end{aligned}
\end{equation}\end{widetext}
and
\begin{widetext}
\begin{equation}
\begin{aligned}\label{H3Eq3a}
& \delta(p_1+p_2-p_3-p_4) u_{p_1}u_{p_2}u_{p_3}v_{p_4}\hat{b}_{p_1}^\dagger \hat{b}_{p_2}^\dagger  \hat{b}_{p_3}  \hat{b}_{-p_4}^\dagger \
 \to\ \delta(p_1-p_2-p_3-p_4) u_{p_1}u_{p_2}u_{p_3}v_{p_4}\hat{b}_{p_3}^\dagger \hat{b}_{p_2}^\dagger  \hat{b}_{p_1}  \hat{b}_{p_4}^\dagger  \\
 = &\ \delta(p_1-p_2-p_3-p_4) u_{p_1}u_{p_2}u_{p_3}v_{p_4}\Big[\hat{b}_{p_4}^\dagger \hat{b}_{p_3}^\dagger \hat{b}_{p_2}^\dagger  \hat{b}_{p_1} +\delta(p_1-p_4)\hat{b}_{p_3}^\dagger \hat{b}_{p_2}^\dagger \Big]\\
\to &\ \delta(p_1-p_2-p_3-p_4) u_{p_1}u_{p_2}v_{p_3}u_{p_4}\hat{b}_{p_4}^\dagger \hat{b}_{p_3}^\dagger \hat{b}_{p_2}^\dagger  \hat{b}_{p_1} \
+\  u_{p_1}^2u_{p_2}v_{p_2} \hat{b}_{p_1}^\dagger \hat{b}_{-p_1}^\dagger, \\
      ~~ & ~~ \\ 
& \delta(p_1+p_2-p_3-p_4) v_{p_1}u_{p_2}v_{p_3}v_{p_4}\hat{b}_{-p_1} \hat{b}_{p_2}^\dagger  \hat{b}_{-p_3}^\dagger  \hat{b}_{-p_4}^\dagger \
\to\ \delta(p_1-p_2-p_3-p_4)v_{p_1}u_{p_2}v_{p_3}v_{p_4}\hat{b}_{p_1} \hat{b}_{p_2}^\dagger  \hat{b}_{p_3}^\dagger  \hat{b}_{p_4}^\dagger \\
 = &\ \delta(p_1-p_2-p_3-p_4)v_{p_1}u_{p_2}v_{p_3}v_{p_4}\Big[  \hat{b}_{p_4}^\dagger \hat{b}_{p_3}^\dagger  \hat{b}_{p_2}^\dagger \hat{b}_{p_1} + \delta(p_1-p_2) \hat{b}_{p_3}^\dagger \hat{b}_{p_4}^\dagger\
+ \ \delta(p_1-p_3)\hat{b}_{p_2}^\dagger \hat{b}_{p_4}^\dagger + \delta(p_1-p_4)\hat{b}_{p_2}^\dagger \hat{b}_{p_3}^\dagger\Big] \\
\to & \ \delta(p_1-p_2-p_3-p_4)v_{p_1}v_{p_2}u_{p_3}v_{p_4}  \hat{b}_{p_4}^\dagger \hat{b}_{p_3}^\dagger  \hat{b}_{p_2}^\dagger \hat{b}_{p_1}\
+ \ \hat{b}^\dagger_{p_1}\hat{b}^\dagger_{-p_1}[v_{p_1}^2u_{p_2}v_{p_2}+2u_{p_1}v_{p_1}v_{p_2}^2],\\
                 ~~ & ~~ \\ 
&\delta(p_1+p_2-p_3-p_4)v_{p_1}v_{p_2}u_{p_3}v_{p_4}\hat{b}_{-p_1} \hat{b}_{-p_2}  \hat{b}_{p_3}  \hat{b}_{-p_4}^\dagger \
 \to\ \delta(p_1-p_2-p_3-p_4) v_{p_1}v_{p_2}u_{p_3}v_{p_4}\hat{b}_{p_4} \hat{b}_{p_2}  \hat{b}_{p_3}  \hat{b}_{p_1}^\dagger\\
= &\  \delta(p_1-p_2-p_3-p_4) v_{p_1}v_{p_2}u_{p_3}v_{p_4}\Big[\hat{b}_{p_1}^\dagger \hat{b}_{p_2}  \hat{b}_{p_3}\hat{b}_{p_4} + \delta(p_1-p_2) \hat{b}_{p_3}\hat{b}_{p_4}+  \delta(p_1-p_3) \hat{b}_{p_2}\hat{b}_{p_4} +  \delta(p_1-p_4) \hat{b}_{p_2}\hat{b}_{p_3} \Big]\\
& \to \delta(p_1-p_2-p_3-p_4) v_{p_1}v_{p_2}u_{p_3}v_{p_4}\hat{b}_{p_1}^\dagger \hat{b}_{p_2}  \hat{b}_{p_3}\hat{b}_{p_4} +  \hat{b}_{p_1}\hat{b}_{-p_1}[v_{p_1}^2u_{p_2}v_{p_2}+2u_{p_1}v_{p_1}v_{p_2}^2]\\
                ~~ & ~~ \\ 
& \delta(p_1+p_2-p_3-p_4) u_{p_1}v_{p_2}v_{p_3}v_{p_4}\hat{b}_{p_1}^\dagger \hat{b}_{-p_2}  \hat{b}_{-p_3}^\dagger  \hat{b}_{-p_4}^\dagger \
\to \ \delta(p_1-p_2-p_3-p_4)v_{p_1}u_{p_2}v_{p_3}v_{p_4}\hat{b}_{p_2}^\dagger \hat{b}_{p_1}  \hat{b}_{p_3}^\dagger  \hat{b}_{p_4}^\dagger\\
 =& \ \delta(p_1-p_2-p_3-p_4)v_{p_1}u_{p_2}v_{p_3}v_{p_4}\Big[\hat{b}_{p_4}^\dagger \hat{b}_{p_3}^\dagger \hat{b}_{p_2}^\dagger \hat{b}_{p_1} + \delta(p_1-p_3)\hat{b}_{p_2}^\dagger \hat{b}_{p_4}^\dagger + \delta(p_1-p_4)\hat{b}_{p_2}^\dagger \hat{b}_{p_3}^\dagger \Big]\\
 \to & \  \delta(p_1-p_2-p_3-p_4)v_{p_1}v_{p_2}u_{p_3}v_{p_4}    \hat{b}_{p_4}^\dagger \hat{b}_{p_3}^\dagger \hat{b}_{p_2}^\dagger \hat{b}_{p_1}+ 2 \hat{b}_{p_1}^\dagger \hat{b}_{-p_1}^\dagger u_{p_1}v_{p_1}v_{p_2}^2. 
\end{aligned}
\end{equation}\end{widetext}
We finally perform the change of variables
\begin{widetext}
\begin{equation}
\begin{aligned}\label{H3Eq4}
& \delta(p_1+p_2-p_3-p_4)\Big[u_{p_1}u_{p_2}v_{p_3}v_{p_4}\hat{b}_{p_1}^\dagger \hat{b}_{p_2}^\dagger \hat{b}_{-p_3}^\dagger \hat{b}_{-p_4}^\dagger+v_{p_1}v_{p_2}u_{p_3}u_{p_4} \hat{b}_{-p_1} \hat{b}_{-p_2}\hat{b}_{p_3}\hat{b}_{p_4}\Big]\\
\to & \ \delta(p_1+p_2+p_3+p_4)u_{p_1}u_{p_2}v_{p_3}v_{p_4}\Big[\hat{b}_{p_1}^\dagger \hat{b}_{p_2}^\dagger \hat{b}_{p_3}^\dagger \hat{b}_{p_4}^\dagger+\hat{b}_{p_1} \hat{b}_{p_2}\hat{b}_{p_3}\hat{b}_{p_4}\Big].
\end{aligned}
\end{equation}\end{widetext}
Combining \eqref{H3Eq1}-\eqref{H3Eq4}, we find the new form \eqref{H3NewForm} for $\hat{H}_3$.

\begin{thebibliography}{10}


\bibitem{Nordheim}
L.W. Nordheim.
\newblock Transport phenomena in {E}instein-{B}ose and {F}ermi-{D}irac gases.
\newblock {\em Proc. Roy. Soc. London A}, 119:689, 1928.

\bibitem{bogoliubov1947theory}
N.~Bogoliubov.
\newblock On the theory of superfluidity.
\newblock {\em J. Phys}, 11(1):23, 1947.


\bibitem{PomeauBinh}
Y.~Pomeau and M.-B. Tran.
\newblock {\em Statistical Physics of Non Equilibrium Quantum Phenomena},
\newblock Lecture Notes in Physics Volume 967, Springer: Berlin/Heidelberg, 2019.



\bibitem{ReichlGust:2012:CII}
E.~D. Gust and L.~E. Reichl.
\newblock Collision integrals in the kinetic equations of dilute bose-einstein
  condensates.
\newblock {\em arXiv:1202.3418}, 2012.








\bibitem{KD1}
T.~R. Kirkpatrick and J.~R. Dorfman.
\newblock Transport theory for a weakly interacting condensed {B}ose gas.
\newblock {\em Phys. Rev. A (3)}, 28(4):2576--2579, 1983.

\bibitem{KD2}
T.~R. Kirkpatrick and J.~R. Dorfman.
\newblock Transport in a dilute but condensed nonideal bose gas: Kinetic
  equations.
\newblock {\em J. Low Temp. Phys.}, 58:301--331, 1985.

\bibitem{akhiezer2013methods}
A.~I. Akhiezer and S.~V. Peletminskii.
\newblock {\em Methods of Statistical Physics}, volume 104 of {\em
  International Series in Natural Philosophy}.
\newblock Elsevier, 1981.

\bibitem{PeletminskiiYatsenko:1968:FTB}
S.~Peletminskii and A.~Yatsenko.
\newblock Contribution to the quantum theory of kinetic and relaxation process.
\newblock {\em Soviet Physics JETP}, 26(773), 1968.

\bibitem{ZarembaNikuniGriffin:1999:DOT}
E.~Zaremba, T.~Nikuni, and A.~Griffin.
\newblock Dynamics of trapped bose gases at finite temperatures.
\newblock {\em J. Low Temp. Phys.}, 116:277--345, 1999.


\bibitem{PomeauBrachetMetensRica}
S.~M'etens Y.~Pomeau, M.A.~Brachet and S.~Rica.
\newblock Th\'eorie cin\'etique d'un gaz de bose dilu\'e avec condensat.
\newblock {\em C. R. Acad. Sci. Paris S'er. IIb M'ec. Phys. Astr.},
  327:791--798, 1999.



\bibitem{QK1}
C.~Gardiner and P.~Zoller.
\newblock Quantum kinetic theory. {A} quantum kinetic master equation for
  condensation of a weakly interacting {B}ose gas without a trapping potential.
\newblock {\em Phys. Rev. A}, 55:2902, 1997.

\bibitem{QK2}
D.~Jaksch, C.~Gardiner, and P.~Zoller.
\newblock Quantum kinetic theory. {II}. {S}imulation of the quantum {B}oltzmann
  master equation.
\newblock {\em Phys. Rev. A}, 56:575, 1997.

\bibitem{QK3}
C.~Gardiner and P.~Zoller.
\newblock Quantum kinetic theory. {III}. {Q}uantum kinetic master equation for
  strongly condensed trapped systems.
\newblock {\em Phys. Rev. A}, 58:536, 1998.

\bibitem{GriffinNikuniZaremba:BCG:2009}
A.~Griffin, T.~Nikuni, and E.~Zaremba.
\newblock Dynamics of trapped bose gases at finite temperatures.
\newblock Cambridge University Press, Cambridge, 2009.






\bibitem{reichl2013transport}
L.~E. Reichl and E.~D. Gust.
\newblock Transport theory for a dilute {B}ose-{E}instein condensate.
\newblock {\em Physical Review A}, 88(5):053603, 2013.








\end{thebibliography}
\subsection{Appendix C}
Let us first compute
\begin{widetext}
\begin{equation}
\begin{aligned}\label{C12Eq1}
& [\hat{b}_{p}^\dagger \hat{b}_{p},\hat{b}_{p_1}^\dagger \hat{b}_{p_2}\hat{b}_{p_3}]
  \ = \  \hat{b}_{p}^\dagger \hat{b}_{p}\hat{b}_{p_1}^\dagger \hat{b}_{p_2}\hat{b}_{p_3}-\hat{b}_{p_1}^\dagger \hat{b}_{p_2}\hat{b}_{p_3}\hat{b}_{p}^\dagger \hat{b}_{p} \ = \  (\delta(p_1'-p_1)-\delta(p-p_2)-\delta(p-p_3))\hat{b}_{p_1}^\dagger \hat{b}_{p_2}\hat{b}_{p_3}.
\end{aligned}
\end{equation}\end{widetext}
We now perform the computation $[\hat{b}_{p_3'}^\dagger \hat{b}_{p_2'}^\dagger \hat{b}_{p_1'}, [\hat{b}_{p}^\dagger \hat{b}_{p},\hat{b}_{p_1}^\dagger \hat{b}_{p_2}\hat{b}_{p_3}]]$. To this end, we compute
\begin{widetext}
\begin{equation}
\begin{aligned}\label{C12Eq2}
& [\hat{b}_{p_3'}^\dagger \hat{b}_{p_2'}^\dagger \hat{b}_{p_1'},\hat{b}^\dagger_{p_1}\hat{b}_{p_2}\hat{b}_{p_3}]
= \ \hat{b}_{p_3'}^\dagger \hat{b}_{p_2'}^\dagger \hat{b}_{p_1'}\hat{b}^\dagger_{p_1}\hat{b}_{p_2}\hat{b}_{p_3} - \hat{b}^\dagger_{p_1}\hat{b}_{p_2}\hat{b}_{p_3}\hat{b}_{p_3'}^\dagger \hat{b}_{p_2'}^\dagger \hat{b}_{p_1'}\\
& =  \delta(p_1'-p_1)\hat{b}_{p_3'}^\dagger \hat{b}_{p_2'}^\dagger \hat{b}_{p_2}\hat{b}_{p_3} -\delta(p_3-p_2')\hat{b}_{p_1}^\dagger \hat{b}_{p_3'}^\dagger \hat{b}_{p_2}\hat{b}_{p_1'}  -\delta(p_3-p_3')\hat{b}_{p_1}^\dagger \hat{b}_{p_2'}^\dagger \hat{b}_{p_2}\hat{b}_{p_1'}\\
&  -\delta(p_2-p_2')\hat{b}_{p_1}^\dagger \hat{b}_{p_3'}^\dagger \hat{b}_{p_3}\hat{b}_{p_1'}-\delta(p_2-p_3')\hat{b}_{p_1}^\dagger \hat{b}_{p_2'}^\dagger \hat{b}_{p_3}\hat{b}_{p_1'}  - \delta(p_2-p_3')\delta(p_3-p_2')\hat{b}_{p_1}^\dagger \hat{b}_{p_1'} - \delta(p_2-p_2')\delta(p_3-p_3')\hat{b}_{p_1}^\dagger \hat{b}_{p_1}.
\end{aligned}
\end{equation}\end{widetext}
Taking into account the fact  $p_1=p_2+p_3$ and $p_1'=p_2'+p_3'$, it now follows straightforwardly from Wick's theorem that
\begin{equation}
\begin{aligned}\label{C12Eq3}& \langle [\hat{b}_{p_3'}^\dagger \hat{b}_{p_2'}^\dagger \hat{b}_{p_1'},\hat{b}^\dagger_{p_1}\hat{b}_{p_2}\hat{b}_{p_3}]\rangle\delta(p_1-p_2-p_3)\delta(p_1'-p_2'-p_3')
 \\
& =  \delta(p_1-p_1')\delta(p_2-p_2')\delta(p_3-p_3')\\
&\times \delta(p_1-p_2-p_3)[f(p_2)f(p_3)-f(p_1)f(p_2)\\
&-f(p_1)f(p_3)-f(p_1)] \\
& +  \delta(p_1-p_1')\delta(p_2-p_3')\delta(p_3-p_2')\\
&\times \delta(p_1-p_2-p_3)[f(p_2)f(p_3)-f(p_1)f(p_2)\\
&-f(p_1)f(p_3)-f(p_1)],
\end{aligned}
\end{equation}
which implies
\begin{equation}
\begin{aligned}\label{C12Eq4}& \langle [\hat{b}_{p_3'}^\dagger \hat{b}_{p_2'}^\dagger \hat{b}_{p_1'}, [\hat{b}_{p}^\dagger \hat{b}_{p},\hat{b}_{p_1}^\dagger \hat{b}_{p_2}\hat{b}_{p_3}]]\rangle\delta(p_1-p_2-p_3)\delta(p_1'-p_2'-p_3')
 \\&=  [\delta(p_1-p_1')\delta(p_2-p_2')\delta(p_3-p_3')\\
 &+\delta(p_1-p_1')\delta(p_2-p_3')\delta(p_3-p_2')]\\
&\times (\delta(p-p_1)-\delta(p-p_2)-\delta(p-p_3))\delta(p_1-p_2-p_3)\\
& \times[f(p_2)f(p_3)-f(p_1)f(p_2)-f(p_1)f(p_3)-f(p_1)].
\end{aligned}
\end{equation}
In the above computation, for  $p_1'=p_1$, there are two choices of $p_2'$ and $p_3'$, $p_2'=p_2$, $p_3'=p_3$ and $p_2'=p_3$, $p_3'=p_2$.

A similar procedure also gives
\begin{equation}
\begin{aligned}\label{C12Eq5}& \langle [\hat{b}_{p_1'}^\dagger \hat{b}_{p_2'}\hat{b}_{p_3'}, [\hat{b}_{p}^\dagger \hat{b}_{p},\hat{b}_{p_3}^\dagger \hat{b}_{p_2}^\dagger \hat{b}_{p_1}]]\rangle\delta(p_1-p_2-p_3)\delta(p_1'-p_2'-p_3')
 \\
 & = [\delta(p_1-p_1')\delta(p_2-p_2')\delta(p_3-p_3')\\
 & +\delta(p_1-p_1')\delta(p_2-p_3')\delta(p_3-p_2')]\\
&\times (\delta(p-p_1)-\delta(p-p_2)-\delta(p-p_3))\delta(p_1-p_2-p_3)\\
&\times[f(p_2)f(p_3)-f(p_1)f(p_2)-f(p_1)f(p_3)-f(p_1)].
\end{aligned}
\end{equation}
Since in the above procedure, the nonlinearity $f(p_2)f(p_3)(f(p_1)+1)-f(p_1)(f(p_2)+1)(f(p_3)+1)$ appears $4$ times, we multiply the factor $\pi {\frac{g^2n}{V}}$ by $4$ and obtain
 the first collision operator $C_{12}$. 
 \subsection{Appendix D}
 We first compute
\begin{equation}
\begin{aligned}\label{C22Eq1}
& [\hat{b}_{p}^\dagger \hat{b}_{p},\hat{b}_{p_1}^\dagger \hat{b}_{p_2}^\dagger \hat{b}_{p_3}\hat{b}_{p_4}] \ = \  \hat{b}_{p}^\dagger \hat{b}_{p}\hat{b}_{p_1}^\dagger \hat{b}_{p_2}^\dagger \hat{b}_{p_3}\hat{b}_{p_4}-\hat{b}_{p_1}^\dagger \hat{b}_{p_2}^\dagger \hat{b}_{p_3}\hat{b}_{p_4}\hat{b}_{p}^\dagger \hat{b}_{p}
 \\
 & \ = \  (\delta(p-p_1)+\delta(p-p_2)-\delta(p-p_3)-\delta(p-p_4))\\
 &\ \ \times \hat{b}_{p_1}^\dagger \hat{b}_{p_2}^\dagger \hat{b}_{p_3}\hat{b}_{p_4}.
\end{aligned}
\end{equation}
We now analyze the commutator $[\hat{b}_{p_1'}^\dagger \hat{b}_{p_2'}^\dagger \hat{b}_{p_3'}\hat{b}_{p_4'}, [\hat{b}_{p}^\dagger \hat{b}_{p},\hat{b}_{p_1}^\dagger \hat{b}_{p_2}^\dagger \hat{b}_{p_3}\hat{b}_{p_4}]]$. To this end, we compute
\begin{equation}
\begin{aligned}\label{C12Eq2}
& [\hat{b}_{p_1'}^\dagger \hat{b}_{p_2'}^\dagger \hat{b}_{p_3'}\hat{b}_{p_4'},\hat{b}^\dagger_{p_1}\hat{b}^\dagger_{p_2}\hat{b}_{p_3}\hat{b}_{p_4}]\\
&= \hat{b}_{p_1'}^\dagger \hat{b}_{p_2'}^\dagger \hat{b}_{p_3'}\hat{b}_{p_4'}\hat{b}^\dagger_{p_1}\hat{b}^\dagger_{p_2}\hat{b}_{p_3}\hat{b}_{p_4}-  \hat{b}^\dagger_{p_1}\hat{b}^\dagger_{p_2}\hat{b}_{p_3}\hat{b}_{p_4} \hat{b}_{p_1'}^\dagger \hat{b}_{p_2'}^\dagger \hat{b}_{p_3'}\hat{b}_{p_4'}\\
& =  \delta(p_3'-p_2)\hat{b}_{p_1'}^\dagger \hat{b}_{p_2'}^\dagger \hat{b}_{p_1}^\dagger \hat{b}_{p_4'}\hat{b}_{p_3}\hat{b}_{p_4}\\
&\  + \  \delta(p_4'-p_2)\hat{b}_{p_1'}^\dagger \hat{b}_{p_2'}^\dagger \hat{b}_{p_1}^\dagger \hat{b}_{p_3'}\hat{b}_{p_3}\hat{b}_{p_4}\\
&\ + \ \delta(p_3'-p_1)\hat{b}_{p_1'}^\dagger \hat{b}_{p_2'}^\dagger \hat{b}_{p_2}^\dagger \hat{b}_{p_4'}\hat{b}_{p_3}\hat{b}_{p_4}\\
&\ + \  \delta(p_4'-p_1)\hat{b}_{p_1'}^\dagger \hat{b}_{p_2'}^\dagger \hat{b}_{p_2}^\dagger \hat{b}_{p_3'}\hat{b}_{p_3}\hat{b}_{p_4}\\
&\ + \ \delta(p_3'-p_1)\delta(p_4'-p_2)\hat{b}_{p_1'}^\dagger \hat{b}_{p_2'}^\dagger \hat{b}_{p_3}\hat{b}_{p_4}\\
&\  + \  \delta(p_4'-p_1)\delta(p_3'-p_2)\hat{b}_{p_1'}^\dagger \hat{b}_{p_2'}^\dagger \hat{b}_{p_3}\hat{b}_{p_4}\\
&\ - \  \delta(p_3-p_2')\hat{b}_{p_1}^\dagger \hat{b}_{p_2}^\dagger \hat{b}_{p_1'}^\dagger \hat{b}_{p_4}\hat{b}_{p_3'}\hat{b}_{p_4'}\\
& \ - \  \delta(p_4-p_2')\hat{b}_{p_1}^\dagger \hat{b}_{p_2}^\dagger \hat{b}_{p_1'}^\dagger \hat{b}_{p_3}\hat{b}_{p_3'}\hat{b}_{p_4'}\\
&\ - \ \delta(p_3-p_1')\hat{b}_{p_1}^\dagger \hat{b}_{p_2}^\dagger \hat{b}_{p_2'}^\dagger \hat{b}_{p_4}\hat{b}_{p_3'}\hat{b}_{p_4'}\\
&\ - \  \delta(p_4-p_1')\hat{b}_{p_1}^\dagger \hat{b}_{p_2}^\dagger \hat{b}_{p_2'}^\dagger \hat{b}_{p_3}\hat{b}_{p_3'}\hat{b}_{p_4'}\\
&\ - \ \delta(p_3-p_1')\delta(p_4-p_2')\hat{b}_{p_1}^\dagger \hat{b}_{p_2}^\dagger \hat{b}_{p_3'}\hat{b}_{p_4'}\\
&\ - \  \delta(p_4-p_1')\delta(p_3-p_2')\hat{b}_{p_1}^\dagger \hat{b}_{p_2}^\dagger \hat{b}_{p_3'}\hat{b}_{p_4'}.
\end{aligned}
\end{equation}
Our next task is to perform Wick's theorem to the $12$ terms. We only analyze below one of them. The other terms can be done in exactly the same way. We compute, 
\begin{equation}
\begin{aligned}\label{C12Eq3}
&\  \delta(p_4'-p_2)\langle \hat{b}_{p_1'}^\dagger \hat{b}_{p_2'}^\dagger \hat{b}_{p_1}^\dagger \hat{b}_{p_3'}\hat{b}_{p_3}\hat{b}_{p_4} \rangle\\
& = \ \delta(p_4'-p_2)\langle \hat{b}_{p_1'}^\dagger  \hat{b}_{p_3'} \rangle \langle \hat{b}_{p_2'}^\dagger \hat{b}_{p_3} \rangle\langle \hat{b}_{p_1}^\dagger \hat{b}_{p_4} \rangle \\
&\ + \ \delta(p_4'-p_2)\langle \hat{b}_{p_1'}^\dagger  \hat{b}_{p_3'} \rangle \langle \hat{b}_{p_2'}^\dagger \hat{b}_{p_4}  \rangle\langle \hat{b}_{p_1}^\dagger \hat{b}_{p_3} \rangle \\
&\ + \  \delta(p_4'-p_2)\langle \hat{b}_{p_1'}^\dagger  \hat{b}_{p_3} \rangle \langle \hat{b}_{p_2'}^\dagger \hat{b}_{p_3'} \rangle\langle \hat{b}_{p_1}^\dagger \hat{b}_{p_4} \rangle\\
&\ +\  \delta(p_4'-p_2)\langle \hat{b}_{p_1'}^\dagger  \hat{b}_{p_3} \rangle \langle \hat{b}_{p_2'}^\dagger \hat{b}_{p_4}  \rangle\langle \hat{b}_{p_1}^\dagger \hat{b}_{p_3'} \rangle\\
&\ + \  \delta(p_4'-p_2)\langle \hat{b}_{p_1'}^\dagger  \hat{b}_{p_4} \rangle \langle \hat{b}_{p_2'}^\dagger \hat{b}_{p_3'} \rangle\langle \hat{b}_{p_1}^\dagger \hat{b}_{p_3} \rangle\\
&\ +\  \delta(p_4'-p_2)\langle \hat{b}_{p_1'}^\dagger  \hat{b}_{p_4} \rangle \langle \hat{b}_{p_2'}^\dagger \hat{b}_{p_3}  \rangle\langle \hat{b}_{p_1}^\dagger \hat{b}_{p_3'} \rangle\\
&=  \delta(p_4'-p_2)\delta({p_1'}-{p_3'})\delta({p_2'}-{p_3} )\delta({p_1}-{p_4})f(p_1')f(p_2')f(p_1)\\
&+ \delta(p_4'-p_2)\delta({p_1'}-{p_3'})\delta({p_2'}-{p_4}) \delta({p_1}-{p_3}) f(p_1')f(p_2')f(p_1)\\
& +   \delta(p_4'-p_2)\delta({p_1'}-{p_3}) \delta({p_2'}-{p_3'})\delta(p_1-p_4)f(p_1')f(p_2')f(p_1)\\
& + \delta(p_4'-p_2)\delta({p_1'}-{p_3})\delta({p_2'}-{p_4}) \delta({p_1}-{p_3'}) f(p_1')f(p_2')f(p_1)\\
&+ \delta(p_4'-p_2)\delta({p_1'}-{p_4})\delta({p_2'}-{p_3'}) \delta({p_1}-{p_3})f(p_1')f(p_2')f(p_1)\\
&+ \delta(p_4'-p_2)\delta({p_1'}-{p_4}) \delta({p_2'}-{p_3}) \delta({p_1}-{p_3'})f(p_1')f(p_2')f(p_1).
\end{aligned}
\end{equation}
The six terms will be analyzed in the details below, with the notice that $p_1+p_2=p_3+p_4$ and $p_1'+p_2'=p_3'+p_4'$.
\begin{itemize}
\item[(i)] The first term $\delta(p_4'-p_2)\delta({p_1'}-{p_3'})\delta({p_2'}-{p_3} )\delta({p_1}-{p_4})f(p_1')f(p_2')f(p_1)$ appears when $p_4'=p_2'=p_2=p_3$, $p_1'=p_3'$, $p_1=p_4$. This  term will cancel with a similar term coming from $\delta(p_4-p_2')\langle \hat{b}_{p_1}^\dagger \hat{b}_{p_2}^\dagger \hat{b}_{p_1'}^\dagger \hat{b}_{p_3}\hat{b}_{p_3'}\hat{b}_{p_4'} \rangle$.
\item[(ii)] The second term $\delta(p_4'-p_2)\delta({p_1'}-{p_3'})\delta({p_2'}-{p_4} )\delta({p_1}-{p_3})f(p_1')f(p_2')f(p_1)$ appears when $p_4'=p_2'=p_2=p_4$, $p_1'=p_3'$, $p_1=p_3$. This  term will cancel with a similar term coming from $\delta(p_4-p_2')\langle \hat{b}_{p_1}^\dagger \hat{b}_{p_2}^\dagger \hat{b}_{p_1'}^\dagger \hat{b}_{p_3}\hat{b}_{p_3'}\hat{b}_{p_4'} \rangle$.
\item[(iiii)] The fourth term $\delta(p_4'-p_2)\delta({p_1'}-{p_3})\delta({p_2'}-{p_3'} )\delta({p_1}-{p_4})f(p_1')f(p_2')f(p_1)$ appears when $p_4'=p_1'=p_2=p_3$, $p_4=p_1$, $p_2'=p_3'$. This  term will cancel with a similar term coming from $\delta(p_4-p_2')\langle \hat{b}_{p_1}^\dagger \hat{b}_{p_2}^\dagger \hat{b}_{p_1'}^\dagger \hat{b}_{p_3}\hat{b}_{p_3'}\hat{b}_{p_4'} \rangle$.
\item[(iv)] The fourth term $\delta(p_4'-p_2)\delta({p_1'}-{p_3})\delta({p_2'}-{p_4} )\delta({p_1}-{p_3'})f(p_1')f(p_2')f(p_1)$ appears when $p_4'=p_2$, $p_4=p_2'$, $p_1'=p_3$, $p_1=p_3'$. This produces the nonlinearity $\delta(p_1+p_2-p_3-p_4)f(p_1)f(p_3)f(p_4)$. 
\item[(v)] The fourth term $\delta(p_4'-p_2)\delta({p_1'}-{p_4})\delta({p_2'}-{p_3'} )\delta({p_1}-{p_3})f(p_1')f(p_2')f(p_1)$ appears when $p_4'=p_1'=p_2=p_4$, $p_3=p_1$, $p_2'=p_3'$. This  term will cancel with a similar term coming from $\delta(p_4-p_2')\langle \hat{b}_{p_1}^\dagger \hat{b}_{p_2}^\dagger \hat{b}_{p_1'}^\dagger \hat{b}_{p_3}\hat{b}_{p_3'}\hat{b}_{p_4'} \rangle$.
\item[(vi)] The fourth term $\delta(p_4'-p_2)\delta({p_1'}-{p_4})\delta({p_2'}-{p_3} )\delta({p_1}-{p_3'})f(p_1')f(p_2')f(p_1)$ appears when $p_4'=p_2$, $p_3=p_2'$, $p_1'=p_4$, $p_1=p_3'$. This produces the nonlinearity $\delta(p_1+p_2-p_3-p_4)f(p_1)f(p_3)f(p_4)$.  
\end{itemize}
As a result, the quantity $\delta(p_4'-p_2)\langle \hat{b}_{p_1'}^\dagger \hat{b}_{p_2'}^\dagger \hat{b}_{p_1}^\dagger \hat{b}_{p_3'}\hat{b}_{p_3}\hat{b}_{p_4} \rangle$ produces 2 times the nonlinearity $\delta(p_1+p_2-p_3-p_4)f(p_1)f(p_3)f(p_4)$. 

Finally, in the end, the nonlinearity 
$f(p_1)f(p_2)(f(p_3)+1)(f(p_4)+1)-f(p_3)f(p_4)(f(p_2)+1)(f(p_1)+1)$ appears 4 times in the forms 
\begin{equation}
\begin{aligned}
& \delta(p_1+p_2-p_3-p_4)\delta(p_1-p_1')\delta(p_2-p_2')\delta(p_3-p_3')\delta(p_4-p_4')\\
&\indent \times [f(p_1)f(p_2)(f(p_3)+1)(f(p_4)+1)\\
&\indent -f(p_3)f(p_4)(f(p_2)+1)(f(p_1)+1)],\\
&~~~~~\\
& \delta(p_1+p_2-p_3-p_4)\delta(p_1-p_2')\delta(p_1-p_2')\delta(p_3-p_4')\delta(p_4-p_3')\\
&\indent \times [f(p_1)f(p_2)(f(p_3)+1)(f(p_4)+1)\\
&\indent -f(p_3)f(p_4)(f(p_2)+1)(f(p_1)+1)],\\
&~~~~~\\
& \delta(p_1+p_2-p_3-p_4)\delta(p_1-p_1')\delta(p_2-p_2')\delta(p_3-p_4')\delta(p_4-p_3')\\
&\indent \times [f(p_1)f(p_2)(f(p_3)+1)(f(p_4)+1)\\
&\indent -f(p_3)f(p_4)(f(p_2)+1)(f(p_1)+1)],\\
&\mbox{and }~~~~~\\
& \delta(p_1+p_2-p_3-p_4)\delta(p_1-p_2')\delta(p_2-p_1')\delta(p_3-p_3')\delta(p_4-p_4')\\
&\indent \times [f(p_1)f(p_2)(f(p_3)+1)(f(p_4)+1)\\
&\indent -f(p_3)f(p_4)(f(p_2)+1)(f(p_1)+1)].
\end{aligned}
\end{equation}
We then multiply the factor $\frac{g^2\pi}{4V^2}$ by $4$ and obtain the collision operator $C_{22}$.
 \subsection{Appendix E}
Let us first compute
\begin{equation}
\begin{aligned}\label{C31Eq1}
& [\hat{b}_{p}^\dagger \hat{b}_{p},\hat{b}_{p_1}^\dagger \hat{b}_{p_2}\hat{b}_{p_3}\hat{b}_{p_4}]\  =\   \hat{b}_{p}^\dagger \hat{b}_{p}\hat{b}_{p_1}^\dagger \hat{b}_{p_2}\hat{b}_{p_3}\hat{b}_{p_4}-\hat{b}_{p_1}^\dagger \hat{b}_{p_2}\hat{b}_{p_3}\hat{b}_{p_4}\hat{b}_{p}^\dagger \hat{b}_{p}
 \\
 &  =   (\delta(p-p_1)-\delta(p-p_2)-\delta(p-p_3)-\delta(p-p_4))\hat{b}_{p_1}^\dagger \hat{b}_{p_2}\hat{b}_{p_3}\hat{b}_{p_4}.
\end{aligned}
\end{equation}
We now analyze the commutator $[\hat{b}_{p_4'}^\dagger \hat{b}_{p_3'}^\dagger \hat{b}_{p_2'}^\dagger \hat{b}_{p_1'}, [\hat{b}_{p}^\dagger \hat{b}_{p},\hat{b}_{p_1}^\dagger \hat{b}_{p_2}\hat{b}_{p_3}\hat{b}_{p_4}]]$. To this end, we compute
\begin{widetext}
\begin{equation}
\begin{aligned}\label{C31Eq2}
&  [\hat{b}_{p_4'}^\dagger \hat{b}_{p_3'}^\dagger \hat{b}^\dagger_{p_2'}\hat{b}_{p_1'},\hat{b}^\dagger_{p_1}\hat{b}_{p_2}\hat{b}_{p_3}\hat{b}_{p_4}] \ = \  \hat{b}_{p_4'}^\dagger \hat{b}_{p_3'}^\dagger \hat{b}^\dagger_{p_2'}\hat{b}_{p_1'}\hat{b}^\dagger_{p_1}\hat{b}_{p_2}\hat{b}_{p_3}\hat{b}_{p_4}-\hat{b}^\dagger_{p_1}\hat{b}_{p_2}\hat{b}_{p_3}\hat{b}_{p_4}\hat{b}_{p_4'}^\dagger \hat{b}_{p_3'}^\dagger \hat{b}^\dagger_{p_2'}\hat{b}_{p_1'}
 \\
  \ = \  & -\ \Big[\delta(p_4-p_4')\hat{b}_{p_1}^\dagger \hat{b}_{p_3'}^\dagger \hat{b}_{p_2'}^\dagger \hat{b}_{p_2}\hat{b}_{p_3}\hat{b}_{p_1'}  +  \delta(p_4-p_4')\delta(p_3-p_3')\hat{b}_{p_1}^\dagger \hat{b}_{p_2'}^\dagger \hat{b}_{p_2}\hat{b}_{p_1'}+  \delta(p_4-p_4')\delta(p_2-p_2')\hat{b}_{p_1}^\dagger \hat{b}_{p_3'}^\dagger \hat{b}_{p_3}\hat{b}_{p_1'} \\\
 &\indent\indent+ \ \delta(p_4-p_4')\delta(p_2-p_3')\hat{b}_{p_1}^\dagger \hat{b}_{p_2'}^\dagger \hat{b}_{p_3}\hat{b}_{p_1'}+  \delta(p_4-p_4')\delta(p_3-p_2')\hat{b}_{p_1}^\dagger \hat{b}_{p_3'}^\dagger \hat{b}_{p_2}\hat{b}_{p_1'}  +  \delta(p_4-p_4')\delta(p_2-p_3')\delta(p_3-p_2')\hat{b}_{p_1}^\dagger \hat{b}_{p_1}\\
&\indent\indent + \ \delta(p_4-p_4')\delta(p_3-p_3')\delta(p_2-p_2')\hat{b}_{p_1}^\dagger \hat{b}_{p_1} + \  \delta(p_3-p_4')\hat{b}_{p_1}^\dagger \hat{b}_{p_3'}^\dagger \hat{b}_{p_2'}^\dagger \hat{b}_{p_2}\hat{b}_{p_4}\hat{b}_{p_1'}  +  \delta(p_3-p_4')\delta(p_4-p_3')\hat{b}_{p_1}^\dagger \hat{b}_{p_2'}^\dagger \hat{b}_{p_2}\hat{b}_{p_1'}\\
&\indent\indent + \ \delta(p_3-p_4')\delta(p_2-p_2')\hat{b}_{p_1}^\dagger \hat{b}_{p_3'}^\dagger \hat{b}_{p_4}\hat{b}_{p_1'} +  \delta(p_3-p_4')\delta(p_2-p_3')\hat{b}_{p_1}^\dagger \hat{b}_{p_2'}^\dagger \hat{b}_{p_4}\hat{b}_{p_1'} +  \delta(p_3-p_4')\delta(p_4-p_2')\hat{b}_{p_1}^\dagger \hat{b}_{p_3'}^\dagger \hat{b}_{p_2}\hat{b}_{p_1'} \\
&\indent\indent +  \delta(p_3-p_4')\delta(p_2-p_3')\delta(p_4-p_2')\hat{b}_{p_1}^\dagger \hat{b}_{p_1} + \ \delta(p_3-p_4')\delta(p_4-p_3')\delta(p_2-p_2')\hat{b}_{p_1}^\dagger \hat{b}_{p_1}\\
&\indent\indent + \delta(p_2-p_4')\hat{b}_{p_1}^\dagger \hat{b}_{p_3'}^\dagger \hat{b}_{p_2'}^\dagger \hat{b}_{p_4}\hat{b}_{p_3}\hat{b}_{p_1'}  +   \delta(p_2-p_4')\delta(p_3-p_3')\hat{b}_{p_1}^\dagger \hat{b}_{p_2'}^\dagger \hat{b}_{p_4}\hat{b}_{p_1'} +  \delta(p_2-p_4')\delta(p_4-p_2')\hat{b}_{p_1}^\dagger \hat{b}_{p_3'}^\dagger \hat{b}_{p_3}\hat{b}_{p_1'} \\
&\indent\indent +  \delta(p_2-p_4')\delta(p_4-p_3')\hat{b}_{p_1}^\dagger \hat{b}_{p_2'}^\dagger \hat{b}_{p_3}\hat{b}_{p_1'} +  \delta(p_2-p_4')\delta(p_3-p_2')\hat{b}_{p_1}^\dagger \hat{b}_{p_3'}^\dagger \hat{b}_{p_4}\hat{b}_{p_1'}  +  \delta(p_2-p_4')\delta(p_4-p_3')\delta(p_3-p_2')\hat{b}_{p_1}^\dagger \hat{b}_{p_1}\\
&\indent\indent + \delta(p_2-p_4')\delta(p_3-p_3')\delta(p_4-p_2')\hat{b}_{p_1}^\dagger \hat{b}_{p_1} +  \delta(p_2-p_3')\delta(p_3-p_2')\hat{b}_{p_1}^\dagger \hat{b}_{p_4'}^\dagger \hat{b}_{p_4}\hat{b}_{p_1'}  +  \delta(p_2-p_3')\delta(p_4-p_2')\hat{b}_{p_1}^\dagger \hat{b}_{p_4'}^\dagger \hat{b}_{p_3}\hat{b}_{p_1'}\\
&\indent\indent + \delta(p_3-p_3')\delta(p_2-p_2')\hat{b}_{p_1}^\dagger \hat{b}_{p_4'}^\dagger \hat{b}_{p_4}\hat{b}_{p_1'}  +  \delta(p_3-p_3')\delta(p_4-p_2')\hat{b}_{p_1}^\dagger \hat{b}_{p_4'}^\dagger \hat{b}_{p_2}\hat{b}_{p_1'} +  \delta(p_4-p_3')\delta(p_3-p_2')\hat{b}_{p_1}^\dagger \hat{b}_{p_4'}^\dagger \hat{b}_{p_2}\hat{b}_{p_1'} \\
&\indent\indent+ \ \delta(p_4-p_3')\delta(p_2-p_2')\hat{b}_{p_1}^\dagger \hat{b}_{p_4'}^\dagger \hat{b}_{p_3}\hat{b}_{p_1'} +  \delta(p_2-p_3')\hat{b}_{p_1}^\dagger \hat{b}_{p_4'}^\dagger \hat{b}_{p_2'}^\dagger \hat{b}_{p_3}\hat{b}_{p_4}\hat{b}_{p_1'} +  \delta(p_3-p_3')\hat{b}_{p_1}^\dagger \hat{b}_{p_4'}^\dagger \hat{b}_{p_2'}^\dagger \hat{b}_{p_2}\hat{b}_{p_4}\hat{b}_{p_1'}  \\ 
&\indent\indent +  \delta(p_4-p_3')\hat{b}_{p_1}^\dagger \hat{b}_{p_4'}^\dagger \hat{b}_{p_2'}^\dagger \hat{b}_{p_2}\hat{b}_{p_3}\hat{b}_{p_1'}  + \delta(p_2-p_2')\hat{b}_{p_1}^\dagger \hat{b}_{p_4'}^\dagger \hat{b}_{p_3'}^\dagger \hat{b}_{p_3}\hat{b}_{p_4}\hat{b}_{p_1'} +  \delta(p_3-p_2')\hat{b}_{p_1}^\dagger \hat{b}_{p_4'}^\dagger \hat{b}_{p_3'}^\dagger \hat{b}_{p_2}\hat{b}_{p_4}\hat{b}_{p_1'}\\
&\indent\indent + \ \delta(p_4-p_2')\hat{b}_{p_1}^\dagger \hat{b}_{p_4'}^\dagger \hat{b}_{p_3'}^\dagger \hat{b}_{p_2}\hat{b}_{p_3}\hat{b}_{p_1'}\Big]  +  \delta(p_1-p_1')\hat{b}_{p_4'}^\dagger \hat{b}_{p_3'}^\dagger \hat{b}_{p_2'}^\dagger \hat{b}_{p_2}\hat{b}_{p_3}\hat{b}_{p_4}.
\end{aligned}
\end{equation}
\end{widetext}

In the above structure, we can see that there are totally $34$ terms, classified into three categories
\begin{itemize}
\item[(i)] $10$ terms of the type $\hat{b}^\dagger \hat{b}^\dagger \hat{b}^\dagger \hat{b}\hat{b}\hat{b}$. 
\item[(ii)] $18$ terms of the type $\hat{b}^\dagger \hat{b}^\dagger \hat{b}\hat{b}$. 
\item[(iii)] $6$ terms of the type $\hat{b}^\dagger \hat{b}^\dagger \hat{b}\hat{b}$. 
\end{itemize}

We provide below the detailed analysis for all of the $10$ terms in the first category. The treatment of the other terms can be done in similar manners.  By Wick's theorem applied to $\delta(p_1-p_1')\hat{b}_{p_4'}^\dagger \hat{b}_{p_3'}^\dagger \hat{b}_{p_2'}^\dagger \hat{b}_{p_2}\hat{b}_{p_3}\hat{b}_{p_4}$, we have
\begin{widetext}
\begin{equation}
\begin{aligned}\label{C31Eq3}
\delta(p_1'-p_1)\langle \hat{b}_{p_4'}^\dagger \hat{b}_{p_3'}^\dagger \hat{b}_{p_2'}^\dagger \hat{b}_{p_2}\hat{b}_{p_3}\hat{b}_{p_4} \rangle
=  &\  \delta(p_1'-p_1)\langle \hat{b}_{p_4'}^\dagger  \hat{b}_{p_4} \rangle \langle \hat{b}_{p_3'}^\dagger \hat{b}_{p_3} \rangle\langle \hat{b}_{p_2'}^\dagger \hat{b}_{p_2} \rangle + \delta(p_1'-p_1)\langle \hat{b}_{p_4'}^\dagger  \hat{b}_{p_4} \rangle \langle \hat{b}_{p_3'}^\dagger \hat{b}_{p_2}  \rangle\langle \hat{b}_{p_2'}^\dagger \hat{b}_{p_3} \rangle \\
& +   \delta(p_1'-p_1)\langle \hat{b}_{p_4'}^\dagger  \hat{b}_{p_2} \rangle \langle \hat{b}_{p_3'}^\dagger \hat{b}_{p_3} \rangle\langle \hat{b}_{p_2'}^\dagger \hat{b}_{p_4} \rangle  + \delta(p_1'-p_1)\langle \hat{b}_{p_4'}^\dagger  \hat{b}_{p_2} \rangle \langle \hat{b}_{p_3'}^\dagger \hat{b}_{p_4}  \rangle\langle \hat{b}_{p_2'}^\dagger \hat{b}_{p_3} \rangle\\
& +   \delta(p_1'-p_1)\langle \hat{b}_{p_4'}^\dagger  \hat{b}_{p_3} \rangle \langle \hat{b}_{p_2'}^\dagger \hat{b}_{p_2} \rangle\langle \hat{b}_{p_3'}^\dagger \hat{b}_{p_4} \rangle+ \delta(p_1'-p_1)\langle \hat{b}_{p_4'}^\dagger  \hat{b}_{p_3} \rangle \langle \hat{b}_{p_3'}^\dagger \hat{b}_{p_2}  \rangle\langle \hat{b}_{p_2'}^\dagger \hat{b}_{p_4} \rangle\\
= &\ \delta(p_1'-p_1)\delta({p_4'}-{p_4})\delta({p_3'}-{p_3} )\delta({p_2'}-{p_2})f(p_2)f(p_3)f(p_4)\\
&+ \delta(p_1'-p_1)\delta({p_4'}-{p_4})\delta({p_3'}-{p_2}) \delta({p_2'}-{p_3}) f(p_2)f(p_3)f(p_4)\\
& +   \delta(p_1'-p_1)\delta({p_4'}-{p_2}) \delta({p_3'}-{p_3})\delta(p_2'-p_4)f(p_2)f(p_3)f(p_4)\\
& + \delta(p_1'-p_1)\delta({p_4'}-{p_2})\delta({p_3'}-{p_4}) \delta({p_2'}-{p_3}) f(p_2)f(p_3)f(p_4)\\
&+ \delta(p_1'-p_1)\delta({p_4'}-{p_3})\delta({p_2'}-{p_2}) \delta({p_3'}-{p_4})f(p_2)f(p_3)f(p_4)\\
&+ \delta(p_1'-p_1)\delta({p_4'}-{p_3}) \delta({p_3'}-{p_2}) \delta({p_2'}-{p_4})f(p_2)f(p_3)f(p_4).
\end{aligned}
\end{equation}\end{widetext}

Now, similar Wick's theorem arguments can be used for $\delta(p_4-p_4')\hat{b}_{p_1}^\dagger \hat{b}_{p_3'}^\dagger \hat{b}_{p_2'}^\dagger \hat{b}_{p_2}\hat{b}_{p_3}\hat{b}_{p_1'}$. In this case, we get the sum of two terms
\begin{equation}
\label{C31Eq4}\begin{aligned}
&\ \delta(p_1-p_1')\delta(p_2-p_3')\delta(p_3-p_2')\delta(p_4-p_4')f(p_1)f(p_2)f(p_3)\\
& +  \delta(p_1-p_1')\delta(p_2-p_2')\delta(p_3-p_3')\delta(p_4-p_4')f(p_1)f(p_2)f(p_3),
\end{aligned}
\end{equation}
and 
\begin{equation}
\label{C31Eq5}\begin{aligned}
&\ \delta(p_1-p_2)\delta(p_3+p_4)\delta(p_4-p_4') \delta(p_3-p_3')f(p_1)f(p_3)f(p_1')\\
&+  \delta(p_1-p_3)\delta(p_2+p_4)\delta(p_4-p_4') \delta(p_2-p_3')f(p_1)f(p_2)f(p_1'),
\end{aligned}
\end{equation}
where, we have used the fact that $p_1=p_2+p_3+p_4$ and $p_1'=p_2'+p_3'+p_4'$.

Taking the sum with respect to $p_1',p_2',p_3',p_4'$ and $p_1,p_2,p_3,p_4$ the second term \eqref{C31Eq5}, with the kernel $K^{13}_{1,2,3,4}$ we find
\begin{equation}
\label{C31Eq6}\begin{aligned}
&2\sum_{p_1,p_2,p_3,p_4,p_1',p_2',p_3',p_4',\ne 0}K^{13}_{1,2,3,4}K^{13}_{1',2',3',4'}\\
&\times \delta(p_1-p_2-p_3-p_4)\delta(p_1'-p_2'-p_3'-p_4')\delta(p_1-p_3)\\
& \times\delta(p_2+p_4)\delta(p_4-p_4') \delta(p_2-p_3')f(p_1)f(p_2)f(p_1'),
\end{aligned}
\end{equation}
where $K^{13}_{1',2',3',4'}$ is $K^{13}_{1,2,3,4}$, in which $p_1,p_2,p_3,p_4$ are replaced by $p_1',p_2',p_3',p_4'$ and we have used the symmetry of $p_2$ and $p_3$, to get the factor $2$ outside.

The other terms have exactly the same structure and by taking the sum of all of terms like \eqref{C31Eq4} and \eqref{C31Eq6}, we arrive at two big terms
\begin{widetext}
\begin{equation}
\begin{aligned}\label{C31Eq7}
 & \mathfrak{A}_1 \ : = \  \ \delta(p_1'-p_1)\delta({p_4'}-{p_4})\delta({p_3'}-{p_3} )\delta({p_2'}-{p_2})[f(p_1)f(p_3)f(p_4)+f(p_1)f(p_2)f(p_4)+f(p_1)f(p_2)f(p_3)]\\
& + \delta(p_1'-p_1)\delta({p_4'}-{p_4})\delta({p_3'}-{p_2}) \delta({p_2'}-{p_3})[f(p_1)f(p_3)f(p_4)+f(p_1)f(p_2)f(p_4)+f(p_1)f(p_2)f(p_3)]\\
& +   \delta(p_1'-p_1)\delta({p_4'}-{p_2}) \delta({p_3'}-{p_3})\delta(p_2'-p_4)[f(p_1)f(p_3)f(p_4)+f(p_1)f(p_2)f(p_4)+f(p_1)f(p_2)f(p_3)]\\
& + \delta(p_1'-p_1)\delta({p_4'}-{p_2})\delta({p_3'}-{p_4}) \delta({p_2'}-{p_3})[f(p_1)f(p_3)f(p_4)+f(p_1)f(p_2)f(p_4)+f(p_1)f(p_2)f(p_3)]\\
& + \delta(p_1'-p_1)\delta({p_4'}-{p_3})\delta({p_2'}-{p_2}) \delta({p_3'}-{p_4})[f(p_1)f(p_3)f(p_4)+f(p_1)f(p_2)f(p_4)+f(p_1)f(p_2)f(p_3)]\\
& + \delta(p_1'-p_1)\delta({p_4'}-{p_3}) \delta({p_3'}-{p_2}) \delta({p_2'}-{p_4})[f(p_1)f(p_3)f(p_4)+f(p_1)f(p_2)f(p_4)+f(p_1)f(p_2)f(p_3)],
\end{aligned}
\end{equation}
\begin{equation}
\label{C31Eq8}\begin{aligned}
&\mathfrak{A}_2 \ : = \ 12\sum_{p_1,p_2,p_3,p_4,p_1',p_2',p_3',p_4',\ne 0}K^{13}_{1,2,3,4}K^{13}_{1',2',3',4'}\delta(p_1-p_2-p_3-p_4)\delta(p_1'-p_2'-p_3'-p_4')\delta(p_1-p_3)
\\
& \times\delta(p_2+p_4)\delta(p_4-p_4') \delta(p_2-p_3')f(p_1)f(p_2)f(p_1'),
\end{aligned}
\end{equation}\end{widetext}
in which, we have taken into account the symmetry of $p_2,p_3,p_4$, to rearrange the terms and get the factor $12$ in front of the sum in $\mathfrak{A}_2$. This term is indeed negligible due to the delta function $\delta(p-p_1)-\delta(p-p_2)-\delta(p-p_3)-\delta(p-p_4)$ in \eqref{C31Eq1}. To see this, we apply this delta function to the left hand side of \eqref{C31Eq8} and get
\begin{widetext}
\begin{equation}
\label{C31Eq9}\begin{aligned}
& (\delta(p-p_1)-\delta(p-p_2)-\delta(p-p_3)-\delta(p-p_4))\mathfrak{A}_2\\
= & \  (\delta(p-p_1)-\delta(p-p_2)-\delta(p-p_3)-\delta(p-p_4)) 12\sum_{p_1,p_2,p_3,p_4,p_1',p_2',p_3',p_4',\ne 0}K^{13}_{1,2,3,4}K^{13}_{1',2',3',4'}\\
&\ \times\delta(p_1-p_2-p_3-p_4)\delta(p_1'-p_2'-p_3'-p_4')\delta(p_1-p_3)\delta(p_2+p_4)\delta(p_4-p_4') \delta(p_2-p_3')f(p_1)f(p_2)f(p_1')\   =  0.
\end{aligned}
\end{equation}\end{widetext}
The first quantity $\mathfrak{A}_1$ can be combined with \eqref{C31Eq3}, yielding
\begin{widetext}
\begin{equation}
\begin{aligned}\label{C31Eq10}
&\delta(p_1'-p_1)\delta({p_4'}-{p_4})\delta({p_3'}-{p_3} )\delta({p_2'}-{p_2})[f(p_2)f(p_3)f(p_4)-f(p_1)f(p_3)f(p_4)-f(p_1)f(p_2)f(p_4)-f(p_1)f(p_2)f(p_3)]\\
& + \delta(p_1'-p_1)\delta({p_4'}-{p_4})\delta({p_3'}-{p_2}) \delta({p_2'}-{p_3}) [f(p_2)f(p_3)f(p_4)-f(p_1)f(p_3)f(p_4)-f(p_1)f(p_2)f(p_4)-f(p_1)f(p_2)f(p_3)]\\
& +   \delta(p_1'-p_1)\delta({p_4'}-{p_2}) \delta({p_3'}-{p_3})\delta(p_2'-p_4)[f(p_2)f(p_3)f(p_4)-f(p_1)f(p_3)f(p_4)-f(p_1)f(p_2)f(p_4)-f(p_1)f(p_2)f(p_3)]\\
& + \delta(p_1'-p_1)\delta({p_4'}-{p_2})\delta({p_3'}-{p_4}) \delta({p_2'}-{p_3}) [f(p_2)f(p_3)f(p_4)-f(p_1)f(p_3)f(p_4)-f(p_1)f(p_2)f(p_4)-f(p_1)f(p_2)f(p_3)]\\
& + \delta(p_1'-p_1)\delta({p_4'}-{p_3})\delta({p_2'}-{p_2}) \delta({p_3'}-{p_4})[f(p_2)f(p_3)f(p_4)-f(p_1)f(p_3)f(p_4)-f(p_1)f(p_2)f(p_4)-f(p_1)f(p_2)f(p_3)]\\
& + \delta(p_1'-p_1)\delta({p_4'}-{p_3}) \delta({p_3'}-{p_2}) \delta({p_2'}-{p_4})[f(p_2)f(p_3)f(p_4)-f(p_1)f(p_3)f(p_4)-f(p_1)f(p_2)f(p_4)-f(p_1)f(p_2)f(p_3)].
\end{aligned}
\end{equation}\end{widetext}
Notice that in the above procedure, the nonlinearity $[f(p_2)f(p_3)f(p_4)-f(p_1)f(p_3)f(p_4)-f(p_1)f(p_2)f(p_4)-f(p_1)f(p_2)f(p_3)]$ appears $6$ times. 

By similar arguments, applied to terms of the other two categories, we find the full nonlinearity $[f(p_3)f(p_4)f(p_2)(f(p_1)+1)-f(p_1)(f(p_2)+1)(f(p_3)+1)(f(p_4)+1)]$, which also appears $6$ times. Now, due to the commutator  $[\hat{b}_{p_1}^\dagger \hat{b}_{p_2}\hat{b}_{p_3}\hat{b}_{p_4}, [\hat{b}_{p}^\dagger \hat{b}_{p},\hat{b}_{p_4'}^\dagger \hat{b}_{p_3'}^\dagger \hat{b}_{p_2'}^\dagger \hat{b}_{p_1'}]]$, the nonlinearity $[f(p_3)f(p_4)f(p_2)(f(p_1)+1)-f(p_1)(f(p_2)+1)(f(p_3)+1)(f(p_4)+1)]$ appears $12$ times in total. We multiply the factor $\frac{g^2\pi}{4V}$ by $12$ and obtain the third collision operator $C_{31}$. 

\def\cprime{$'$}

\end{document}